\documentclass[twocolumn]{aastex6}

\usepackage{natbib}
\usepackage{graphicx}
\usepackage{amsmath}   
\usepackage{float}
\usepackage[caption=false]{subfig}
\usepackage{wrapfig}
\usepackage{gensymb}
\usepackage{url}
\usepackage{hyperref}   
\usepackage{color}
\usepackage{amssymb}
\usepackage{epstopdf}
\usepackage{calc}

\graphicspath{{./}}
\newcommand{\Msun}{\rm M_{\odot}}

\begin{document}

\title {The nature of Disk of Satellites around Milky Way-like galaxies}

\author
{
Moupiya Maji$^{1, 2}$, Qirong Zhu$^{1,2,3}$, Federico Marinacci$^{4}$,and Yuexing Li$^{1, 2}$}	

\affil{
$^{1}$Department of Astronomy \& Astrophysics, The Pennsylvania State University, University Park, PA 16802, USA \\
$^{2}$Institute for Gravitation and the Cosmos, The Pennsylvania State University, University Park, PA 16802, USA\\
$^{3}$Harvard-Smithsonian Center for Astrophysics, Harvard University, 60 Garden Street, Cambridge, MA 02138, USA\\
$^{4}$Department of Physics, Kavli Institute for Astrophysics and Space Research, Massachusetts Institute of Technology,
Cambridge, MA 02139, USA\\
}

\email{moupiya@astro.psu.edu}

\begin{abstract}

It has been suggested that the satellite galaxies of the Milky Way reside in a highly-flattened, 
kinematically-coherent
plane called Disk of Satellites (DoS). The origin of the DoS, however, has been hotly debated,
and a number of conflicting claims have been reported
in the literature on whether or not the DoS is consistent with predictions from the standard
Lambda Cold Dark Matter ($\Lambda$CDM) cosmological model. Here we
investigate this issue by comparing a high-resolution, hydrodynamic $\Lambda$CDM simulation 
of a Milky Way sized galaxy with its dark matter only
counterpart. We find the following results: (1) The abundance and distribution of satellite galaxies around a host
galaxy is significantly different in the hydro simulation compared to its N-body counterpart;
(2) No clear coherent rotation is
found in the satellite system, as the fractions of corotating and counter-corotating satellites remain
comparable across cosmic time; (3) The satellite distribution 
evolves significantly with time, from nearly isotropic at high redshift to
anisotropic at
the present day;
(4) The DoS properties strongly depend on sample selection and plane identification methods.
Our results imply that the spatially-thin and coherently-rotating DoS reported in Milky Way and other galaxies 
may be a selection effect of small sample size.
\end{abstract}


\maketitle

\section{Introduction}
In the '70s  \cite{Lynden-Bell1976} and \cite{Kunkel1976} found that the 11 brightest satellite galaxies of the
Milky Way (MW) have a highly 
anisotropic distribution and that they align in a plane inclined to the Galactic stellar disk. Such planar
structure is now commonly referred to as ``disk of satellites'' (DoS, \citealt{Kroupa2005}). To date, 
more than two dozens new dwarf 
galaxies have been detected around
the MW (e.g., \citealt{Helmi2008, Willman2010, McConnachie2012, Koposov2015}). It was reported that 
these new dwarfs
also have an anisotropic distribution and can be interpreted as lying in a disk \citep{Pawlowski2015},
although the new DoS is
thicker and has a higher minor-to-major axis ratio (Maji et al. 2017, in prep).
It was reported that
15 out of the 27 dwarfs around Andromeda,  
detected by Pan-Andromeda Archaeological Survey (PAndAS; \citealt{McConnachie2009}),
also follow an anisotropic planar distribution (e.g., \citealt{Koch2006, 
Metz2007, McConnachie2009, Pawlowski2013, Conn2013, Ibata2013}).

Among the original 11 ``classical'' satellites around Milky Way (MW), 
it was reported that 7 to 9 galaxies preferentially co-orbit in a similar direction
\citep{Pawlowski2013}, which have been interpreted
as coherent motion of the DoS.
\cite{Ibata2013} used line-of-sight (LOS)
velocities to
suggest that 13 out of 15 coplanar satellites of Andromeda are co-rotating. 
Outside of the Local Group, \cite{Ibata2014} used the SDSS catalog and identified 22 galaxies with 
diametrically opposed satellite pairs 
and found that 20 of them have anti-correlated velocities, suggesting that co-planar and co-rotating
satellite galaxies are common in the
Universe.

However, these claims have been rebuffed recently. In Maji et al. 2017 (in prep), we performed a 
comprehensive reanalysis of the observed Milky Way satellites. We found that the DoS 
structure depends strongly on sample size and the plane identification method,  and that only 6 out of the 11 ``classical" dwarfs may 
be considered as corotating, in contrast to previous claims \citep{Pawlowski2013, Pawlowski2015methods}. 
Moreover, \cite{Buck2016} performed 21 cosmological simulations to investigate the kinematics of M31 satellites, and 
they concluded that LOS velocities
are not representative of the 3D velocities of the galaxies themselves. When only LOS velocities are used, the results
can apparently agree with
the observations, but when the full 3D angular momenta of the galaxies are considered no coherent motion can be 
found on the DoS plane.
Furthermore, investigations on SDSS galaxies by \cite{Cautun2015b} 
and \cite{Phillips2015}
found that the excess of pairs of anti-correlated galaxies is very sensitive to sample selection parameters
and sample size, and it is consistent
with random noise corresponding to an under-sampling of the data.

The origin of the anisotropic distribution of satellites has been a hotly debated issue. 
Some early studies suggested that satellite galaxies 
preferentially avoid regions near host galaxies equator plane and tend to cluster near the poles 
\citep{Holmberg1969, Zaritsky1997}, but later observations
showed that this may only be true for certain type of galaxies (e.g., \citealt{Sales2004, 
Brainerd2005, Azzaro2007, Bailin2008, Agustsson2010}). 
In recent years, a number of simulations have been aimed at explaining the DoS. Initially, N-body simulations
were largely unsuccessful to directly 
predict the DoS because they produced an isotropic distribution of dark matter sub-halos around the main 
galaxy in the standard Lambda Cold Dark 
Matter ($\Lambda$CDM) cosmology \citep{Kang2005}. This has been strongly criticized as a failure of
$\Lambda$CDM by some authors \citep{Kroupa2005,
Metz2007, Kroupa2010, Pawlowski2012, Pawlowski2014}.

Recent developments in numerical techniques and computational power
have made it possible to study
the DoS phenomenon in a more realistic manner. \cite{Bahl2014} investigated
the probability of finding satellite planes similar to the DoS around M31 in the Millennium II Simulation, and found that such planes occur frequently.
\cite{Sawala2016} analyzed the APOSTLE simulations, a suite of smoothed particle hydrodynamics (SPH)
simulations of the Local Group, and found that satellite systems form with a wide 
range of spatial anisotropies and it is possible to reproduce the observed DoS of 11 brightest MW satellites. 
\cite{Cautun2015a} analyzed two high resolution N-body cosmological simulations 
(Millennium-II; \citealt{msii2009} and Copernicus Complexio; \citealt{Hellwing2016})
and found that planar distribution of satellites are 
very common and the degree of anisotropy vary from system to system.

The controversies surrounding the DoS stem from three separate issues: (i) the plane detection
method or the definition of plane is not well
specified or uniform across different studies, which results in obtaining different results using the same 
sample; (ii) different sample sizes have been
used in various studies and most of them are 
using very small number of galaxies (11 for MW and 15 for M31), and
(iii) the majority of theoretical studies did not include the effect of baryons in the cosmological simulations which can 
strongly  affect the
distribution and abundance of galaxies.

In order to address these controversies,
we investigate all the three issues mentioned above in this study by analyzing a high-resolution cosmological 
hydrodynamic simulation of a MW-sized galaxy by \cite{Marinacci2014}, and compare it to its DM-only 
counterpart. The dwarf galaxies in the simulations
are identified with a density-based hierarchical algorithm, the Amiga Halo Finder \citep{Knollmann2009, 
Gill2004}. A more detailed description of the simulations is given in \cite{Marinacci2014} and \cite{Zhu2015}.
To perform our analysis
we divide the dwarfs into four different sample sizes as found in observations and
analyze their spatial and kinematic properties. We also adopt two different types of
plane identification methods
used in literature. In addition, we track these satellites
to high redshift in order to understand the nature and origin of the distribution of the satellite system.

This paper is organized as follows. 
In \S~\ref{sec:methods} we describe the numerical techniques used in this investigation, which include the plane identification methods, 
the cosmological simulations and dwarf
galaxy identification 
code used. In \S~\ref{sec:results_pos_z0}, we present the abundance and spatial distribution of satellites
at the present day, by comparing hydrodynamical and N-body simulations.

In \S~\ref{sec:results_vel_z0} we show the kinematic properties
of the satellites at redshift $z = 0$. The evolution of the satellite system is explored in
\S~\ref{sec:evolution}. 
We discuss the various selection effects on the DoS, which include the sample size,
the distance of satellites from central galaxy,
and plane detection methods in \S~\ref{sec:discussions}, and we summarize our 
findings in \S~\ref{sec:summary}.

\section{Methods}
\label{sec:methods}

\subsection{Cosmological $\Lambda$CDM Simulations}

In this paper we use the hydrodynamical cosmological simulation of a Milky Way-sized galaxy (Aq-C-4 halo) by \cite{Marinacci2014} (hereafter referred
to as Hydro simulation), and a dark matter-only run of the same halo from \cite{Zhu2016} (hereafter referred to as DMO simulation) for comparison. The
Hydro simulation was performed with the moving mesh code AREPO \citep{Springel2010}, and 
it models, other than gas dynamics, a set of baryonic processes playing a key role in galaxy formation. The model includes
an effective ISM model describing a two-phase interstellar medium, 
star formation, metal-dependent  cooling, metal enrichment and mass return from stellar
evolution, and both stellar and AGN feedback.

The modeled galaxy has a virial mass\footnote{total mass inside virial radius (234 kpc). Here the virial radius is defined as the radius of the
sphere which encloses an overdensity of 200 with respect to the critical density.}
of $1.59 \times 10^{12}\,\Msun$, similar to that of the Milky Way. 
The mass resolution for baryonic particles is
$5\times 10^4\,\Msun$ and for
DM particles it is $2.7\times 10^5\,\Msun$ in the hydrodynamic simulation.
In the DMO run, the DM particles has the summed mass of the two types of particles in hydro run, namely $3.2\times 10^5\Msun$.


\subsection{Satellite Identification}
\label{sec:sat_identification}

There are two general methods of finding halos in a cosmological field: particle-based and density-based.
In particle based methods, e.g.
Friends-of-friends (FOF)
\citep{Davis1985}, all particles within a specified linking length are considered as a candidate halo. 
However, with this class of methods; if there is a linking bridge between two candidate halos, they
are identified as a single structure and a full recovery of all constituting substructures is not possible.

This is not the case for density-based
algorithm, such as the Amiga Halo Finder (AHF, \citealt{Gill2004, Knollmann2009}) 
that we use in this study for identifying the satellites in the 
simulation. AHF first divides the simulation box into grids, computes the density inside each cell and compares it to a threshold value or background
density. If the cell density is higher than the threshold, 
then AHF divides the cell further and repeats the process recursively until all cell have densities under the
threshold value. In the next stage, AHF starts from the finest cell level and 
marks isolated dense regions as structure candidates. It moves up one 
level, again finding probable cluster candidates 
and links clusters in the coarser level to those in the finer level. 
In this way halo-subhalo-subsubhalo tree is built, which is
useful for tracking the progenitors of satellites at higher redshifts. 

\subsection{Plane Identification Methods}
\label{sec:4methods}
We have used two types of methods for plane identification in our paper: 
Principal Component Analysis (PCA) and Tensor of Inertia (TOI) 
method. The TOI method is used in conjunction with with 3 different weighting functions.
We give a more detailed description of both plane identification techniques in the subsections below.

\subsubsection{Principal Component Analysis (PCA)}

PCA is a common method used for multivariate data analysis in statistics. The goal of PCA in general is to explore linear 
relationships between different variables in the dataset and, thus
simplifying the data by reducing its dimensionality. 
In our specific case we aim
to find out to what extent the 3D distribution of the satellite galaxies can be expressed as 2D, 
i.e. galaxies
lying on a plane (DoS). In this application, the method of PCA can be viewed as fitting an ellipsoid to the data.
The anisotropy of the distribution
can thus be expressed as the ratio of minor axis to the major axis $(c/a)$. If the distribution is perfectly 2D
(i.e. tha data lie on an infinitely thin plane), $c/a = 0$, while
if it is perfectly isotropic, $c/a = 1$.
PCA is an orthogonal linear transformation of the data which finds and projects the data on a new co-ordinate system where the highest data
variance lies on the new x-axis, the second highest variance lies on the new y-axis and so on. The steps of this procedure are shown below.

Let the positions of the dwarfs be denoted by $(x_i, y_i, z_i)$ where $i$ 
runs from 1 to $n$, the number of dwarfs in the system.
First the data is centered by subtracting
the mean of the $x$, $y$ and $z$ positions of the dwarfs from their original co-ordinates. The new position matrix is :

\begin{eqnarray}
\boldsymbol{I^{'}} & = & \left(
\begin{array}{ccc}
x_1^{\prime} & y_1^{\prime} & z_1^{\prime}\\
x_2^{\prime}  & y_2^{\prime}  & z_2^{\prime} \\
\dots & \dots & \dots \\
x_n^{\prime}  & y_n^{\prime}  & z_n^{\prime} 
\end{array}\right)
\end{eqnarray}

where ,
\begin{equation}
x_i^{\prime} = x_i - \frac{1}{n}\sum_1^{n} x_j ;\quad y_i^{\prime} = y_i - \frac{1}{n}\sum_1^{n} y_j ;
\quad z_i^{\prime} = z_i - \frac{1}{n}\sum_1^{n} z_j 
\end{equation}
Now we find the covariance matrix of the new positions given by :

\begin{equation}
   {\boldsymbol I}_{\rm covariance}^{\prime} =   {\boldsymbol I}^{\prime T}{\boldsymbol I}^{\prime} 
\end{equation}

This is a $3 \times 3$ dimensional matrix given by:

\begin{eqnarray}
{{\boldsymbol I}_{\rm covariance}^{\prime}} & = & \sum_{i=1}^{n}\left(
\begin{array}{ccc}
x_i^{\prime 2} & x_i^{\prime}y_i^{\prime} & x_i^{\prime}z_i^{\prime}\\
x_i^{\prime}y_i^{\prime} & y_i^{\prime 2} & y_i^{\prime}z_i^{\prime}\\
x_i^{\prime}z_i^{\prime} & y_i^{\prime}z_i^{\prime} & z_i^{\prime 2}
\end{array}\right)
\end{eqnarray}

We find the eigenvalues, $ \lambda = ( \lambda_1, \lambda_2, \lambda_3)$ and 
eigenvectors $\boldsymbol V = (\boldsymbol V_1, \boldsymbol V_2, \boldsymbol V_3)$,  of this matrix and these eigenvectors form the
orthogonal basis of the
new co-ordinate system (the eigenvector with the highest eigenvalue is new x axis and the one with smallest value is new z axis). 
We now find our
centered data in this new co-ordinate system
\begin{equation}
\boldsymbol I_{\rm new} = \boldsymbol I^{'} \times \boldsymbol V 
\end{equation}

We obtain the 3 axes of the fitted ellipsoid $a, b$ and $c$ by taking the standard deviation ($\sigma$) of the new data:

\begin{equation}
\sigma = \sqrt{ \frac{1}{n} {{\boldsymbol I}_{\rm new}^{T} \times \boldsymbol I}_{\rm new}}
\end{equation}

The off diagonal entries in this multiplied matrix are all zero as the cross correlation between positions are zero now and we are 
left with only 
the diagonal terms : $x_{\rm new}^2, y_{\rm new}^2, z_{\rm new}^2$. So, we have $c = \sqrt{\frac{1}{n} z_{\rm new}^2}$ and similarly
for $a$ and $b$.

\begin{figure*}
\centering
\includegraphics[width=0.45\textwidth]{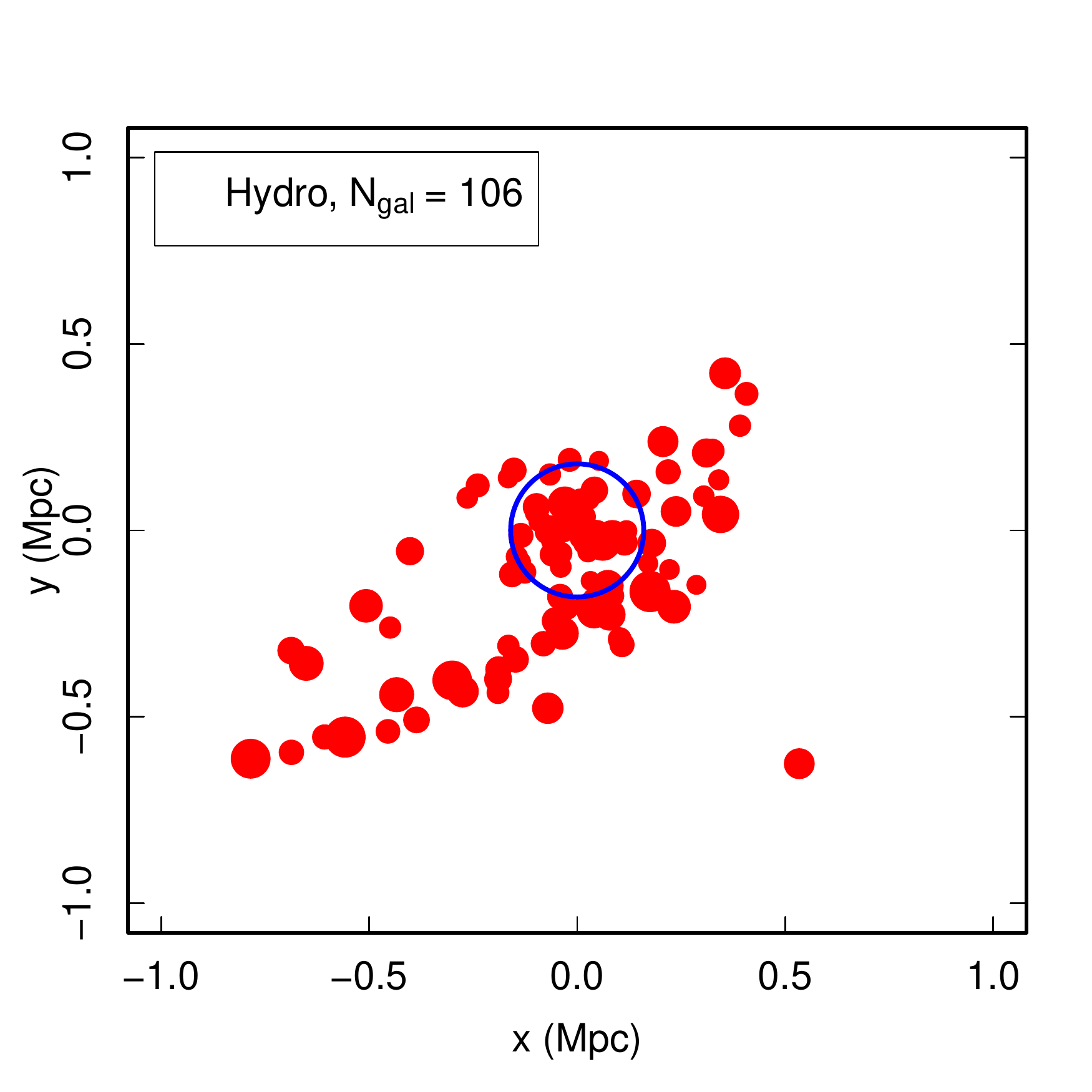} 
\includegraphics[width=0.45\textwidth]{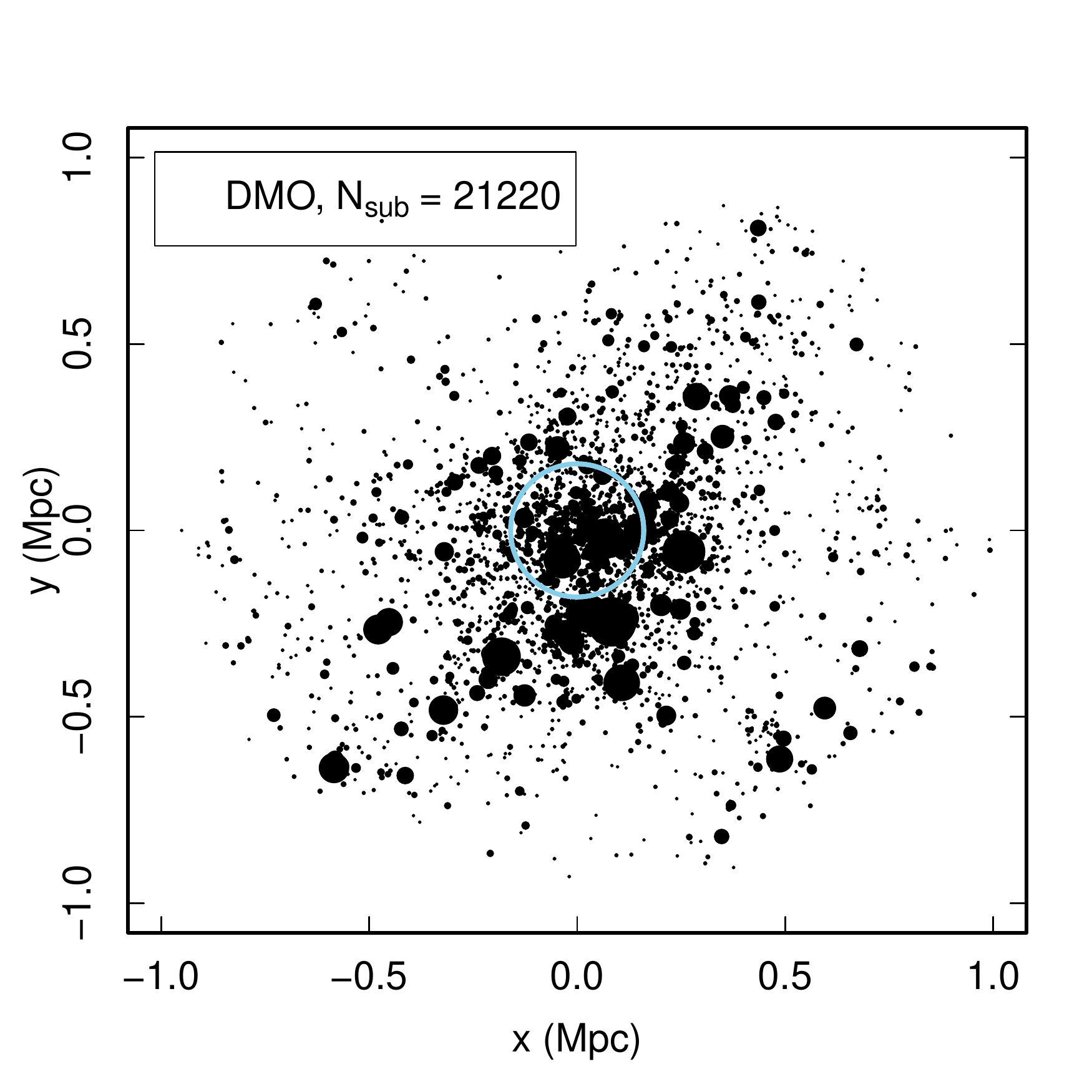} 
\caption{A comparison of the projected positions of dwarfs with total mass above $10^7 \Msun$, 
as represented by filled circles, in the X-Y plane at z=0 within 1 Mpc from
the central
 galaxy between the hydrodynamical (left) and 
the dark matter-only (right) simulations. The size of the circle is proportional to the halo mass, and
the open circle in the center of the two plots indicates the virial radius of the simulated Milky Way at $\sim$ 240 kpc at z=0.
}
\label{fig:twopositions}
\end{figure*}

\subsubsection{Tensor of Inertia method} 

This is a common method that has been used in literature \citep{Pawlowski2015methods} to determine the $a$, $b$, and $c$ axes of satellite
distribution. In this method we find the moment of inertia of the dwarfs and diagonalize it to find the axes length ($a, b, c$) of the distribution.
Let the position vector of a dwarf be given $\boldsymbol r_i = (x_i,y_i, z_i)$. We calculate the tensor of inertia (TOI) matrix, weighted or
unweighted and find the eigenvalues of that matrix. The TOI is a $3 \times 3$ matrix which is given by 

\begin{eqnarray}
{{\boldsymbol I}_{\rm covariance}} & = & \sum_{i=1}^{n}\left(
\begin{array}{ccc}
x_i^2 & x_iy_i & x_iz_i\\
x_iy_i & y_i^2 & y_iz_i\\
x_iz_i & y_iz_i & z_i^2
\end{array}\right)
\times w_i
\end{eqnarray}

This matrix has the same form as the covariance matrix discussed in PCA method. The term $w_i$ in the above equation refers to the weight 
assigned to the dwarfs
in the system according to their distances.
If $w_i= 1$, then this reduces to the standard method where all dwarfs have the same weight
irrespective of their distances. It has been argued that the far off dwarfs should carry less weight in determining the DoS plane as they have higher
chances of being outliers \citep{Bailin2005}. 
For completeness, we use two cases of reduced weights, $w_i = 1/r_i$ and
$w_i = 1/r_i^2$ where $r_i= \sqrt{x_i^2 + y_i^2 +z_i^2}$.
The three eigenvalues of this matrix ${\boldsymbol I}_{\rm TOI}$, gives the three axes of the dwarf distribution $a,b,c$ (largest eigenvalue is a 
and the smallest one is c). The anisotropy of the distribution is characterized by the ratio of the minor-to-major axis, $c/a$ 
(for isotropic distribution $c/a = 1$).
We will use the two methods, PCA and reduced TOI with weights 1, $1/r$ and $1/r^2$ for determining the $c/a$ ratios of our various samples in this paper.
This will enable us to get a description of the dwarf distribution and examine the effects of the different methods discussed 
above.

\section{Abundance and Spatial Distribution of Satellites at z=0}
\label{sec:results_pos_z0}

\subsection{Effects of Baryons}
\label{sec:effect_baryons}

Early $N$-body simulations of galaxies generally showed
a rather isotropic distribution
of dark matter subhalos surrounding the central galaxy \citep{Kang2005}, which has often been interpreted
as a failure of 
$\Lambda$CDM cosmology \citep{Kroupa2005, Metz2007, Kroupa2010, Pawlowski2012, Pawlowski2014}. However,
these simulations did not include the effect of baryons. Recent studies 
have shown that baryons play an important role in determining the properties of both subhalos and the main host
(e.g. \citealt{Zhu2016, Sawala2016}), but little is known about how baryons affect DoS structure.
In this study, we investigate the effect of baryons on the abundance and distribution of satellites
of a MW-like galaxy, by comparing the baryonic simulation and its N-body counterpart 
(details in $\S \ref{sec:4methods}$).

Figure~\ref{fig:twopositions} shows the projected spatial distribution of all subhalos found 
within 1 Mpc of the MW center in both simulations.
We find that there are many more subhalos ($\sim$ 21220) within the DMO simulation compared to the
106 luminous subhalos (dwarf galaxies) in the baryonic run, because
most subhalos in the DMO simulation do not form stars
and hence they are not identified as galaxies.
In \cite{Zhu2016}, we identified three major baryonic processes, namely adiabatic contraction, 
tidal disruption, and reionization, that significantly affect the density distribution of dark matter halos and their ability to retain gas, 
thus reducing
the star formation activity in many low-mass halos.
Moreover, the DMO dwarfs are distributed noticeably more isotropically compared to baryonic dwarfs, which have a clear anisotropic distribution as a 
result of reduced abundance of star-forming dwarfs and interactions with the central galaxy.
Similar results of anisotropic distribution in baryonic simulations were also reported by \cite{Sawala2016}.
Even early studies \citep{Zentner2005} showed that the N-body subhalos are not fully isotropic and the likely luminous subhalos
(satellites) are even more anisotropically distributed.
This figure demonstrates that baryonic processes have a profound impact on the 
abundances and spatial distribution of satellite galaxies around a central
MW-type galaxy, and it illustrates the 
difference between N-body simulations and hydrodynamical simulations on the study of the DoS 
phenomenon (e.g. \citealt{Pawlowski2015methods, Sawala2016}).
Hereafter, we are going to focus on the detailed analysis of the dwarfs in the baryonic simulation.

\begin{figure}
\includegraphics[width=0.49\textwidth]{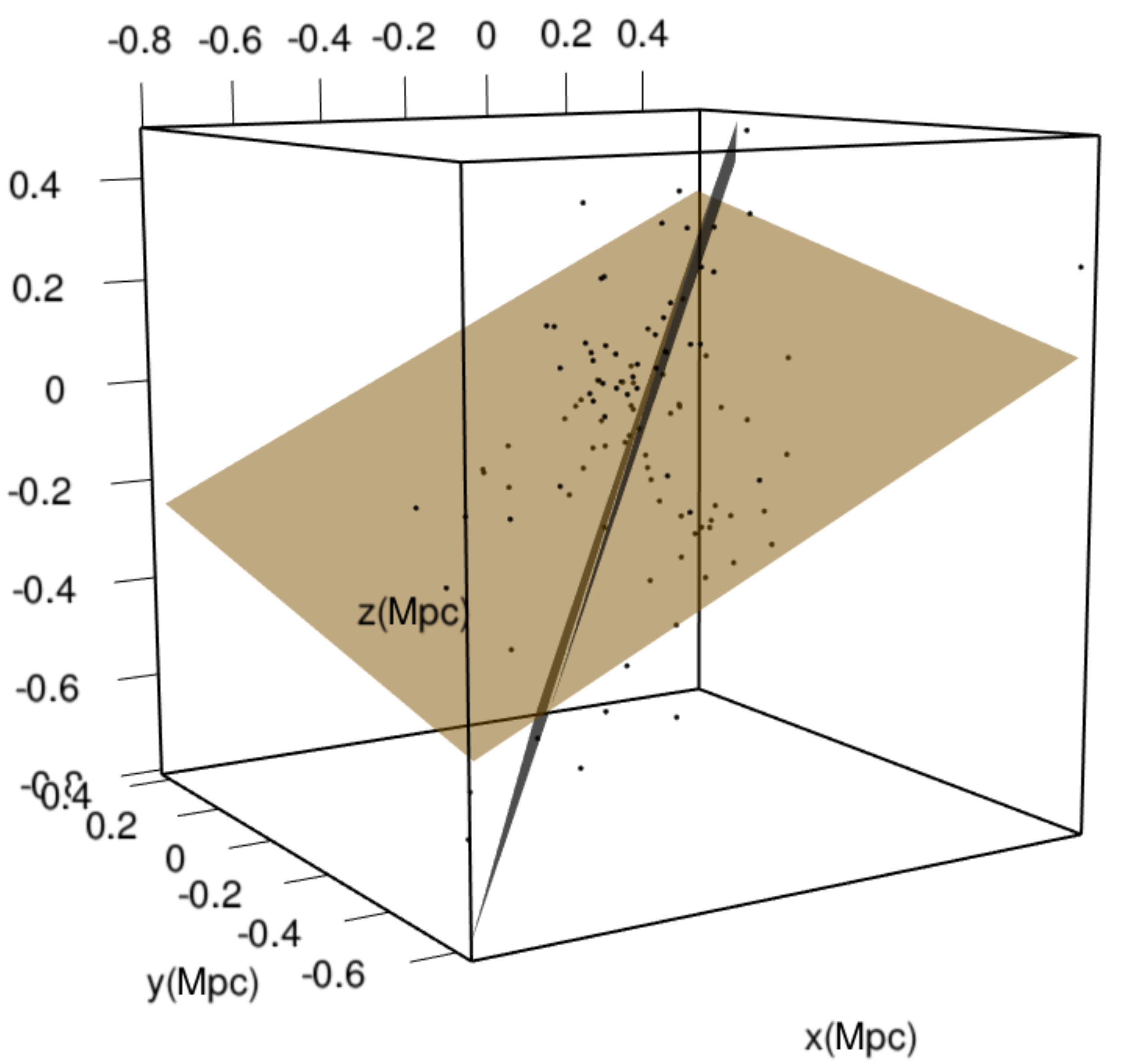} 
\caption{Three-dimensional spatial distribution of satellites (black dots) within 1 Mpc from the central galaxy in the hydrodynamical simulation. 
The fitted plane of the satellites, using PCA method, is shown as the black plane and the disk plane of the 
simulated Milky Way is depicted as the brown plane.}
\label{fig:twoplanes}
\end{figure}

In Figure~\ref{fig:twoplanes} we fit the distribution of
106 dwarfs within 1 Mpc of the central galaxy with the PCA method (see \S~\ref{sec:4methods}). 
Among the 106 dwarfs, about $77\%$ can be considered as residing in the same plane (dwarfs within the rms height),
as indicated by the blue plane in the figure.
 The angle between our fitted DoS plane and the simulated MW disk (as indicated by the orange plane) is $\sim 75$ degrees, which
is very close to the observed angle of $\sim 77.3$ degrees \citep{Pawlowski2013}. 

\begin{figure}
\includegraphics[width=0.49\textwidth]{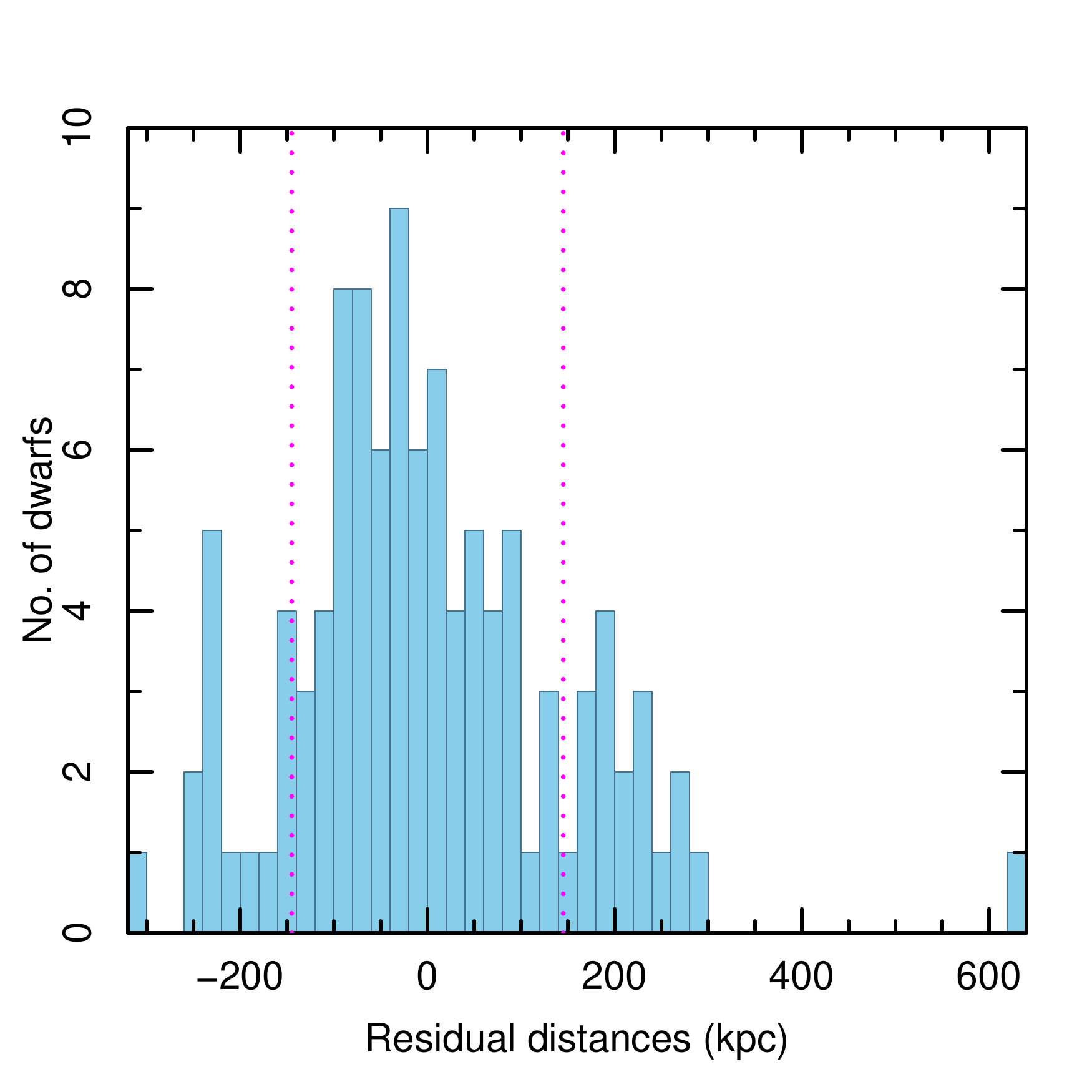}
\caption{Distribution of the residual distances of the simulated satellites from the fitted
DoS plane. 
The width between the two magenta dotted lines shows the root-mean-square height of the plane, 
$\sim 145$ kpc.}
\label{fig:hist_dist_sim}
\end{figure}

However, as shown in Figure~\ref{fig:hist_dist_sim}, the r.m.s. height of the DoS, fitted to 106 dwarfs within 1 Mpc, is $\sim$ 145 kpc, which is 
much larger than those reported from observations ($\sim 30$ kpc for 39 dwarfs within 365 kpc)
which typically use a much smaller sample (\citealt{Pawlowski2015}, Maji et al. 2017 (in prep)). 
We will address the effect of sample size on the DoS properties in the next section.

\subsection{Effects of Sample Size and Plane Identification Method}
\label{sec:spatial}

In Maji et al. 2017 (in prep), we reanalyzed all dwarfs currently detected around the MW and grouped them in three subsets,
as used by different groups in the literature: the 11 'classical' dwarfs 
\citep{Kroupa2005}, the 27 most massive nearby dwarfs and the complete sample of 39 dwarfs \citep{Pawlowski2015}. We fit the three samples using 
both PCA and TOI methods (\ref{sec:4methods}) and found that both the isotropy (as indicated by $c/a$ ratio) and the thickness (characterized 
by the root-mean-square or r.m.s. height) of the DoS 
increase with sample size: the $c/a$ goes from  $\sim 0.2$ for the 11 dwarfs to $\sim 0.26$ for the 39 dwarfs, and the r.m.s. height goes 
from $\sim 20$ kpc to $\sim 30$ kpc.

\begin{figure}
\includegraphics[width=0.49\textwidth]{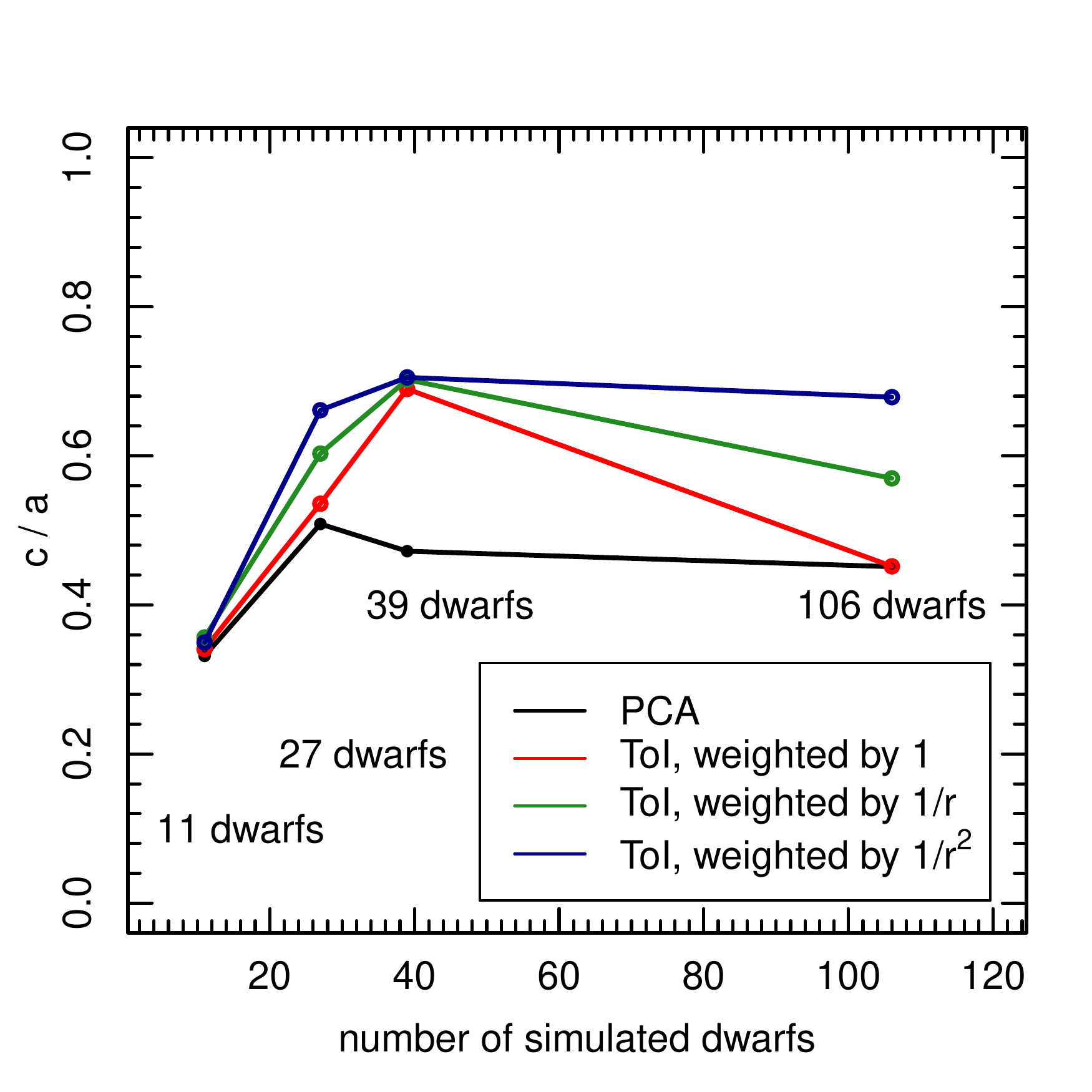}
\caption{The ``isotropy " of the simulated satellite distribution, as indicated by $c/a$, as a function of the sample size.
We have used four samples here with 11, 27, 39 and 106 dwarf galaxies respectively.
The different colors denote the different plane fitting methods : Principal Component Analysis (black) and
Tensor of Inertia with three types of weight functions, 1 (red), $1/r$ (green) and $1/r^2$ (blue).}
\label{fig:ca_sim}
\end{figure}

To directly compare our simulated DoS with observations, 
we first calculate the farthest distance of the dwarfs from MW center for these three subsets and find them to be 257.4 kpc (Leo I), 257.4 kpc (Leo I) 
and 365 kpc (Eri II) for the sample of 11, 27 and 39 dwarfs respectively.
Then we divide our simulated dwarfs into 4 subsets: 11 most massive (total mass) dwarfs
within 257.4 kpc, 27 most massive dwarfs within 257.4 kpc, 39 most massive dwarfs within 365 kpc and for completeness, all 106 dwarfs within 1 Mpc.
We apply the 4 methods discussed in \S~\ref{sec:methods} and determine the $c/a$ ratio for these four samples, as shown in Figure
~\ref{fig:ca_sim}.
For the 11 dwarfs, the c/a ratio is $\sim 0.3$ in our simulation, close to the value of $\sim 0.2$ for the observations \citep{Metz2007, Pawlowski2015}.
However, with the full
sample of 106 simulated dwarfs ($<1$ Mpc), the ratio increases to $\sim 0.44$ for PCA method, and it goes upto 0.68 for ToI method weighted by $1/r^2$.
Similar range of $c/a$ was also reported by \cite{Sawala2016}. This plot demonstrates that the DoS isotropy is subject to selection effect.

\begin{figure}
\includegraphics[width=0.49\textwidth]{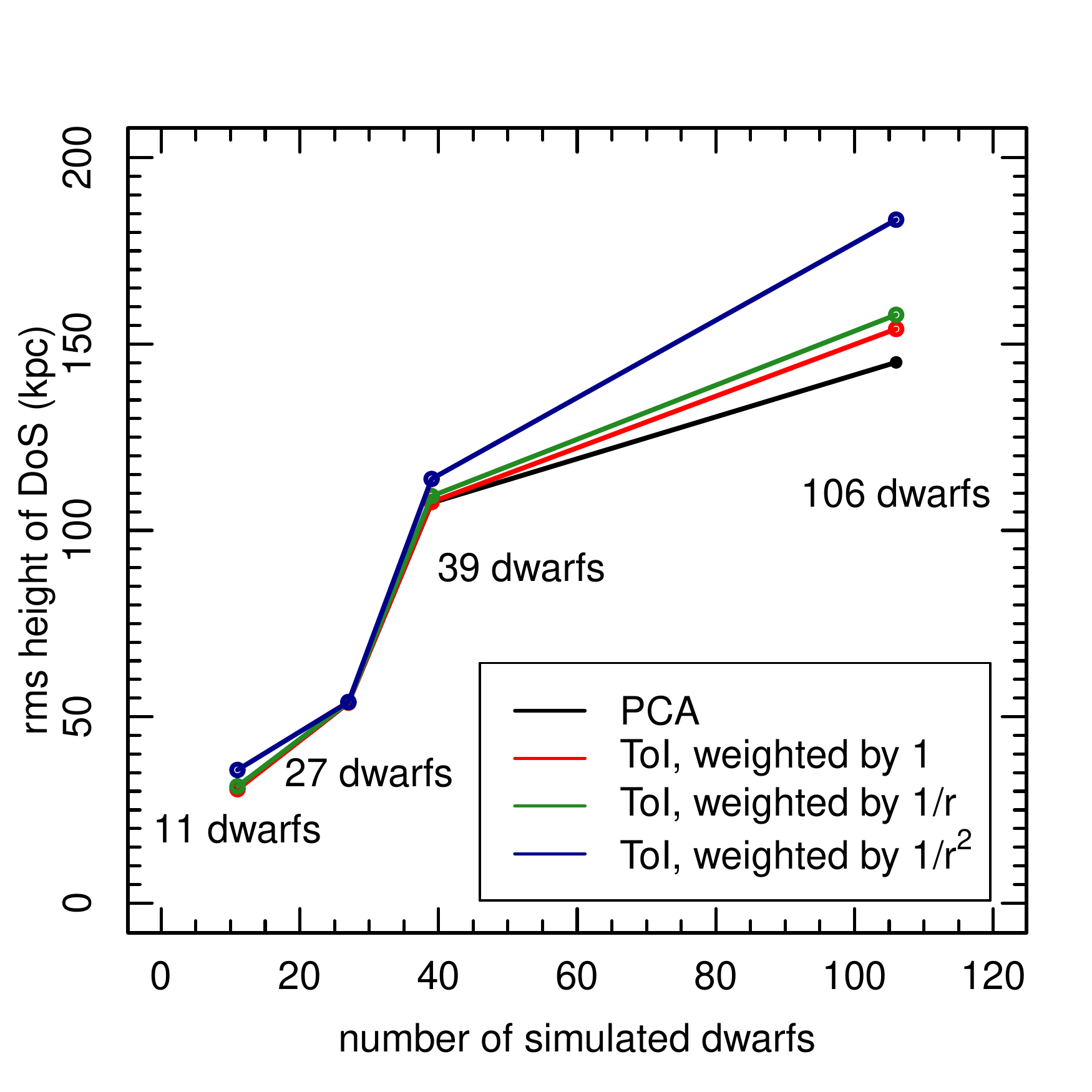}
\caption{The root-mean-square (rms) height of the DoS plane fitted to a sample of simulated dwarfs as a function of the sample size.
We have used the same four samples here with 11, 27, 39 and 106 dwarf galaxies respectively.
The different colors denote same four plane fitting methods : Principal Component Analysis (black) and
Tensor of Inertia with three types of weight functions, 1 (red), $1/r$ (green) and $1/r^2$ (blue).}
\label{fig:rmsh_sim}
\end{figure}

Furthermore,  in Figure~\ref{fig:rmsh_sim} 
we show the r.m.s. heights of the simulated DoS plane for the four samples using four different methods
and find that the DoS plane height increases with sample size.
Using PCA method, the r.m.s. height for 11 dwarfs
is $\sim 30$ kpc, and it rises to 120 kpc for 39 dwarfs and to 145 kpc when we include all dwarfs out to 1 Mpc. 
The thickness of the fitted DoS is even larger when ToI methods are used, as shown in the figure.
For comparison,
when \cite{Kroupa2005} took into account all known observed dwarfs upto 1 Mpc, the calculated r.m.s. DoS height was 159 kpc, close to our simulated
height.

These results demonstrate that the DoS properties change significantly with the sample size of the satellites and the plane 
identification method. We have seen similar trends of changing DoS properties in observed satellites too as discussed in
Maji et al. 2017, in prep. These considerations suggest that the properties of the highly flattened DoS of the MW
derived from a small set of observed 
dwarfs \citep{Ibata2013, Pawlowski2015methods} may not be robust .

\section{Kinematic Properties of Satellites at z=0}
\label{sec:results_vel_z0}

\begin{figure}
 \includegraphics[width=0.5\textwidth]{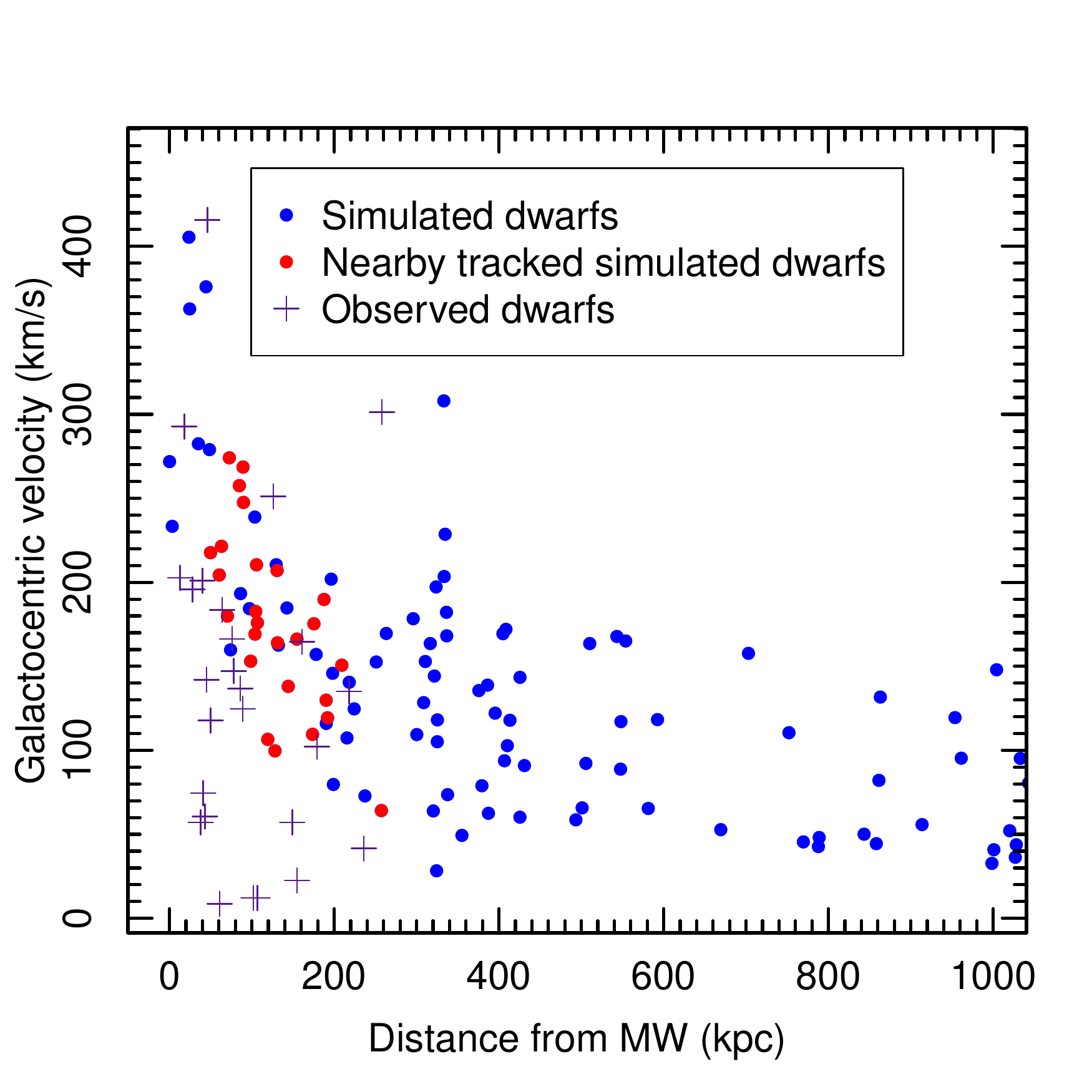}
 \caption{Comparison of galactocentric velocity between the simulated (blue points) and observed (purple crosses) dwarfs as a function of 
 galactocentric distance. The observed data is taken from \cite{McConnachie2012}, and velocities are calculated with respect to the 
 Galaxy. The red points are the 27 most massive dwarfs within 257.4 kpc of galactic center from the simulation.}
\label{fig:veldist}
\end{figure}

Another claim of the DoS is that the satellites have coherent rotation in the same plane. It was suggested by \cite{Pawlowski2013} that, of
the 11 ``classical" dwarfs that have proper motion measurements, 7 to 9 are corotating on the DoS. However, in the analysis of 
Maji et al. 2017 (in prep.) 
we found that only 6 meet  the criterion of corotation, and no firm coherent motion can be inferred. Moreover,
these studies suffer from a 
very large uncertainties in the velocity measurements and a relatively small sample size.
A larger sample size and precise measurements of 3-D velocity 
components of dwarfs from our simulation provide an advantageous study of the kinematic properties of the satellites. 

Figure~\ref{fig:veldist} shows the galactocentric velocity of the simulated dwarfs within 1 Mpc from the central galaxy, compared to
Milky Way observations \cite{McConnachie2012}. Most of the observed dwarfs are located within 300~kpc from the MW, and they have a velocity range 
from $\sim 10$ to $\sim 400$~km/s. While the simulated ones have a similar velocity range, they have a higher median velocity ($\sim 150$ 
km/s compared to $\sim 100$ km/s from the observations).

\begin{figure}
 \includegraphics[width=0.49\textwidth]{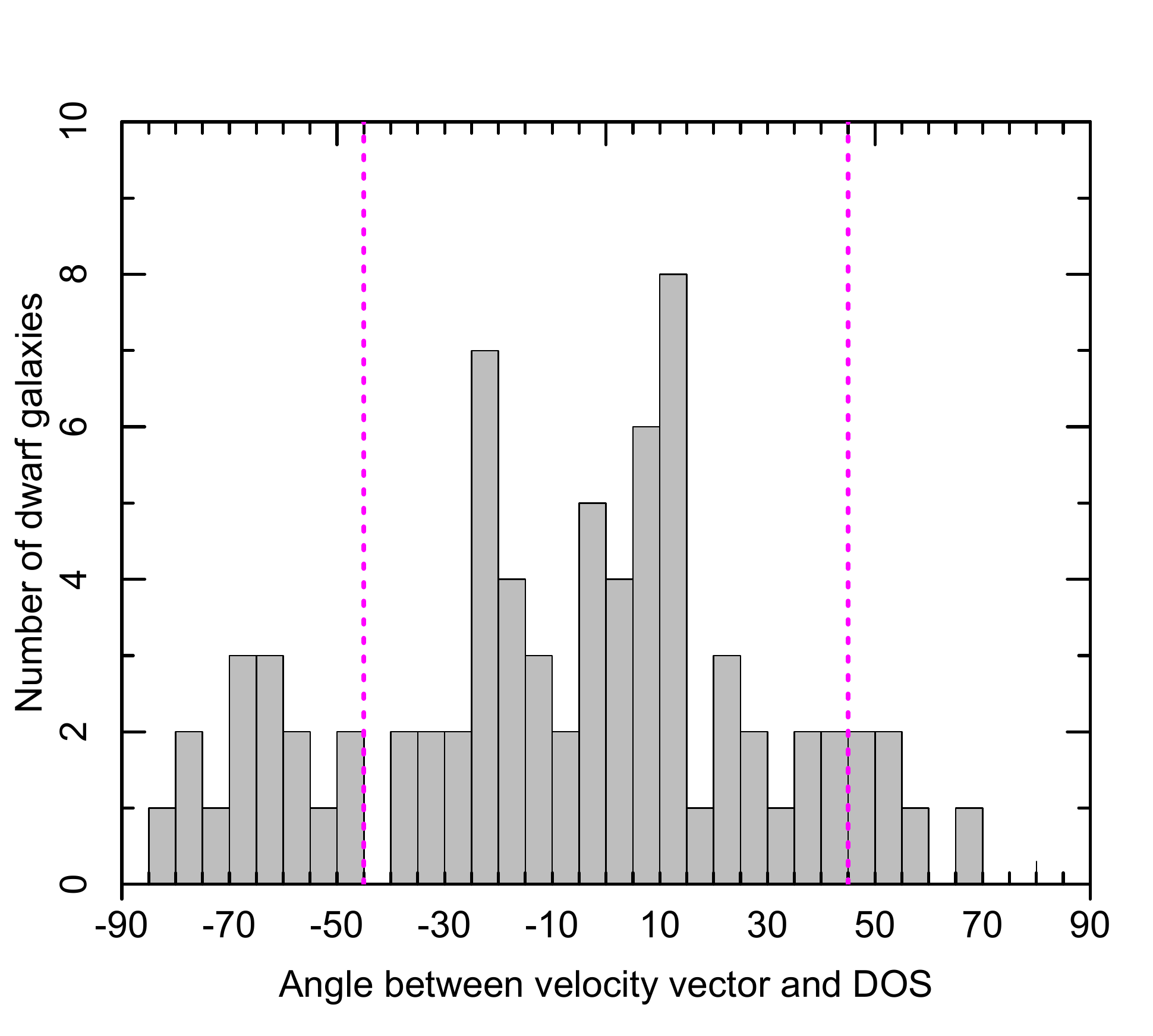}
 \caption{Distribution of the angle between velocity vector of simulate dwarfs and the DoS plane. The dwarfs whose velocity 
 lies within -45 degree to +45 
 degree (magenta dotted lines) of the DoS are considered to be moving on the DoS.}
\label{fig:hist_vv_dos} 
\end{figure}

\begin{figure}
 \includegraphics[width=0.49\textwidth]{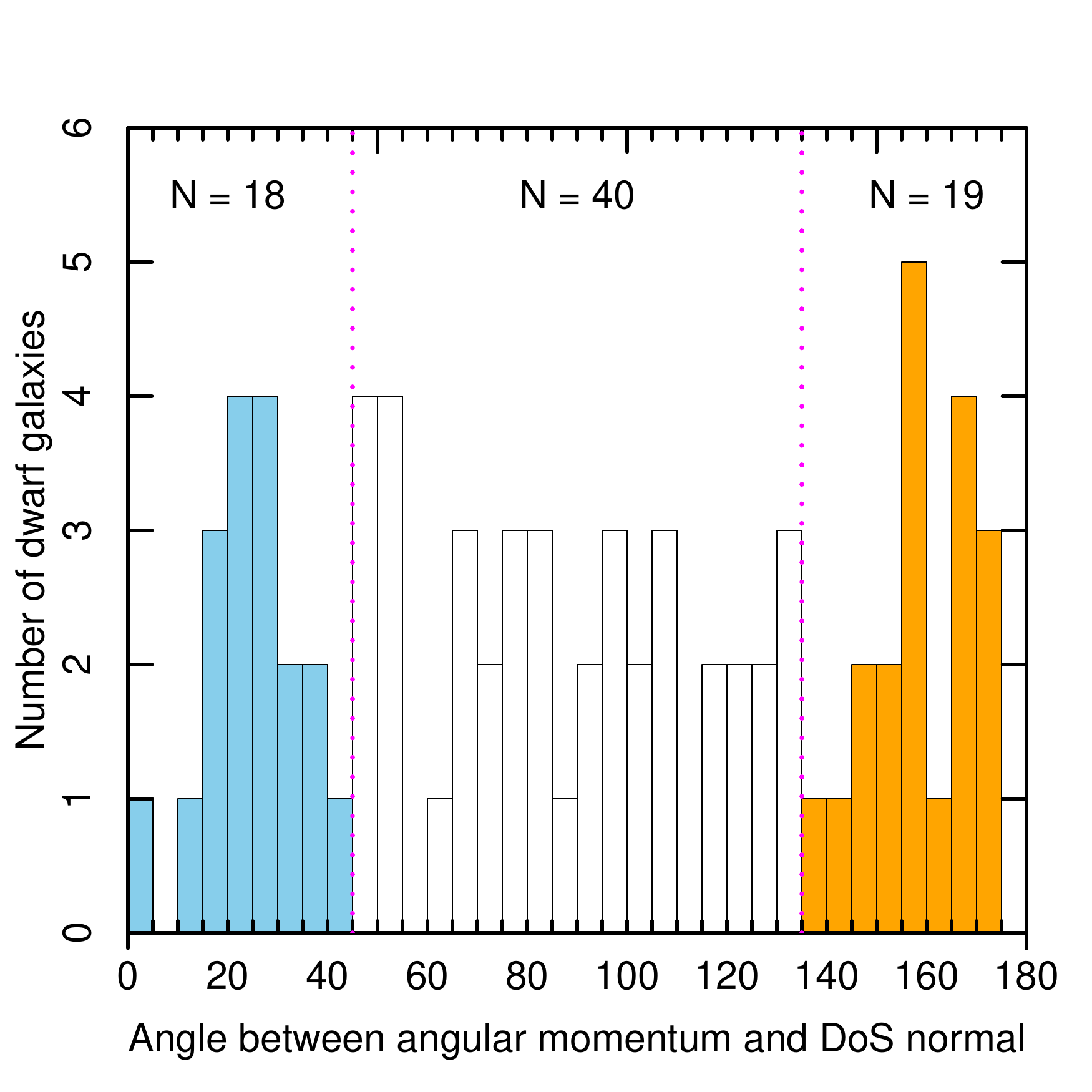}
 \caption{Distribution of angle between angular momentum of simulated dwarfs (residing in the DoS plane) and the DoS normal. 
 There are 18 corotating (blue) and 19 counter-corotating (orange) dwarfs.}
 \label{fig:hist_l_ndos}
\end{figure}

To investigate the kinematic coherence of the DoS, we first calculate the fraction of dwarfs moving in the DoS plane. We use a criterion to define a
satellite as moving on the DoS when it total velocity vector falls  from -45 to +45 degree of the DoS plane. As shown in Figure~\ref{fig:hist_vv_dos},
77 out of 106,  or 73$\%$ of dwarfs within 1 Mpc, meet this criterion and are considered to be moving on the DoS.  
These dwarfs can be moving either mostly circularly (clockwise or counter clockwise) or mostly radially on the DoS.

Next, we calculate the fraction of dwarfs in the same circular motions (either corotation or counter-corotation) to determine whether or not the DoS 
is rotationally supported. We use a criterion to define a satellite as rotating on the DoS when its angular momentum vector falls from 0 to 45 degree
of the DoS normal, or counter-corotating when the angle between the two is from 135 degree to 180 degree. As shown in Figure~\ref{fig:hist_l_ndos}, out
of the 77 satellites moving on the DoS, 18 are corotating and 19 are counter-corotating. If we consider all 106 satellites within 1 Mpc from the
simulation, only $17\%$ are corotating and 18$\%$ are counter-corotating. These numbers strongly argue against any trend of corotation or
counter-corotation, instead they show that the DoS is not rotationally supported.

\begin{figure}
\includegraphics[width=0.49\textwidth]{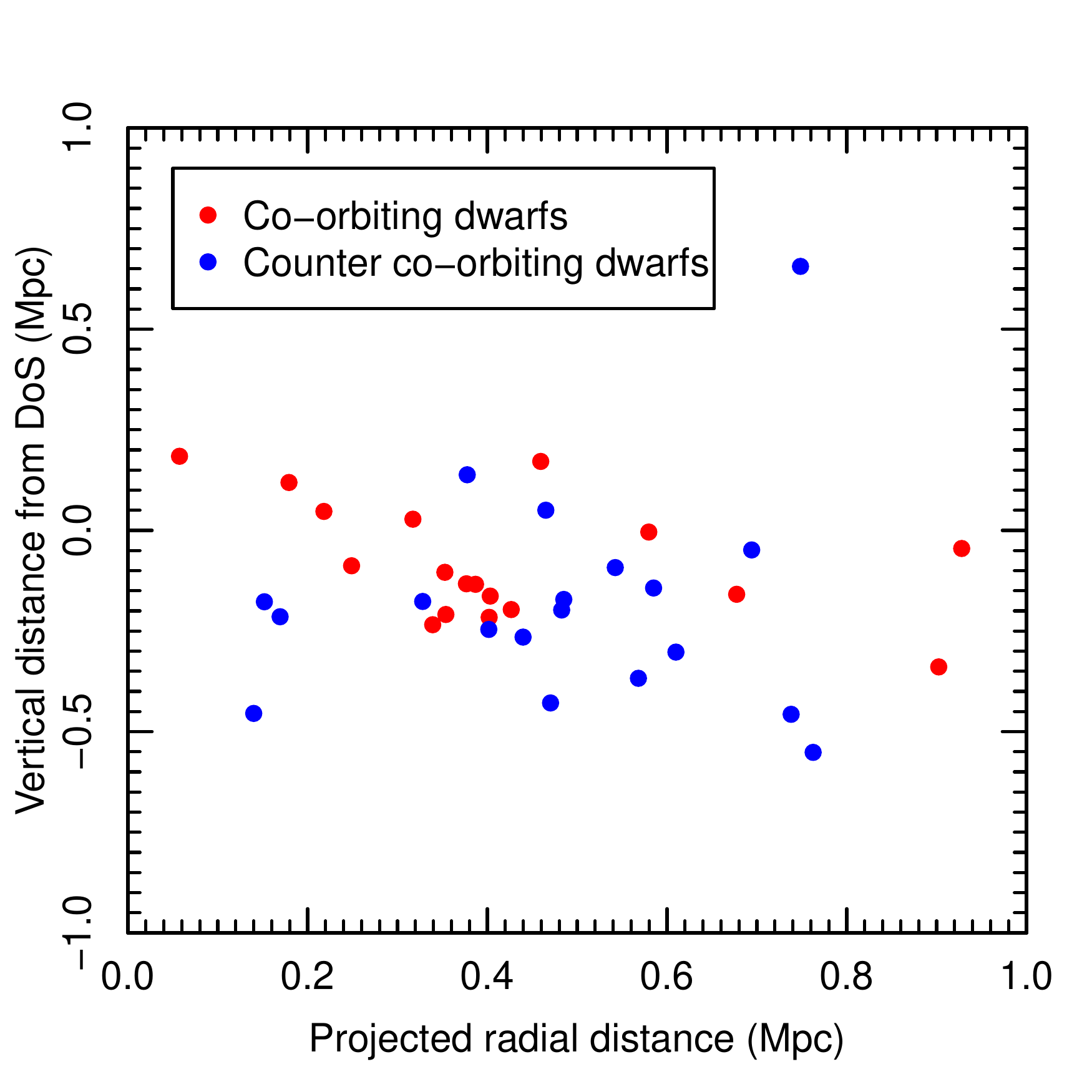} 
 \caption{Vertical distance of the simulated dwarfs from the DoS vs their 
projected radial distance (i.e. the distance component lying on the DoS) from the center of the 
 DoS (center-of-mass of
 the satellite galaxies). The red points show corotating dwarfs and blue points show counter-corotating dwarfs.}
 \label{fig:vert_rad_dist}
\end{figure}

\begin{figure*}
 \centering
 \includegraphics[width=\textwidth,clip=true,trim=0pt 0pt 0pt 0pt]{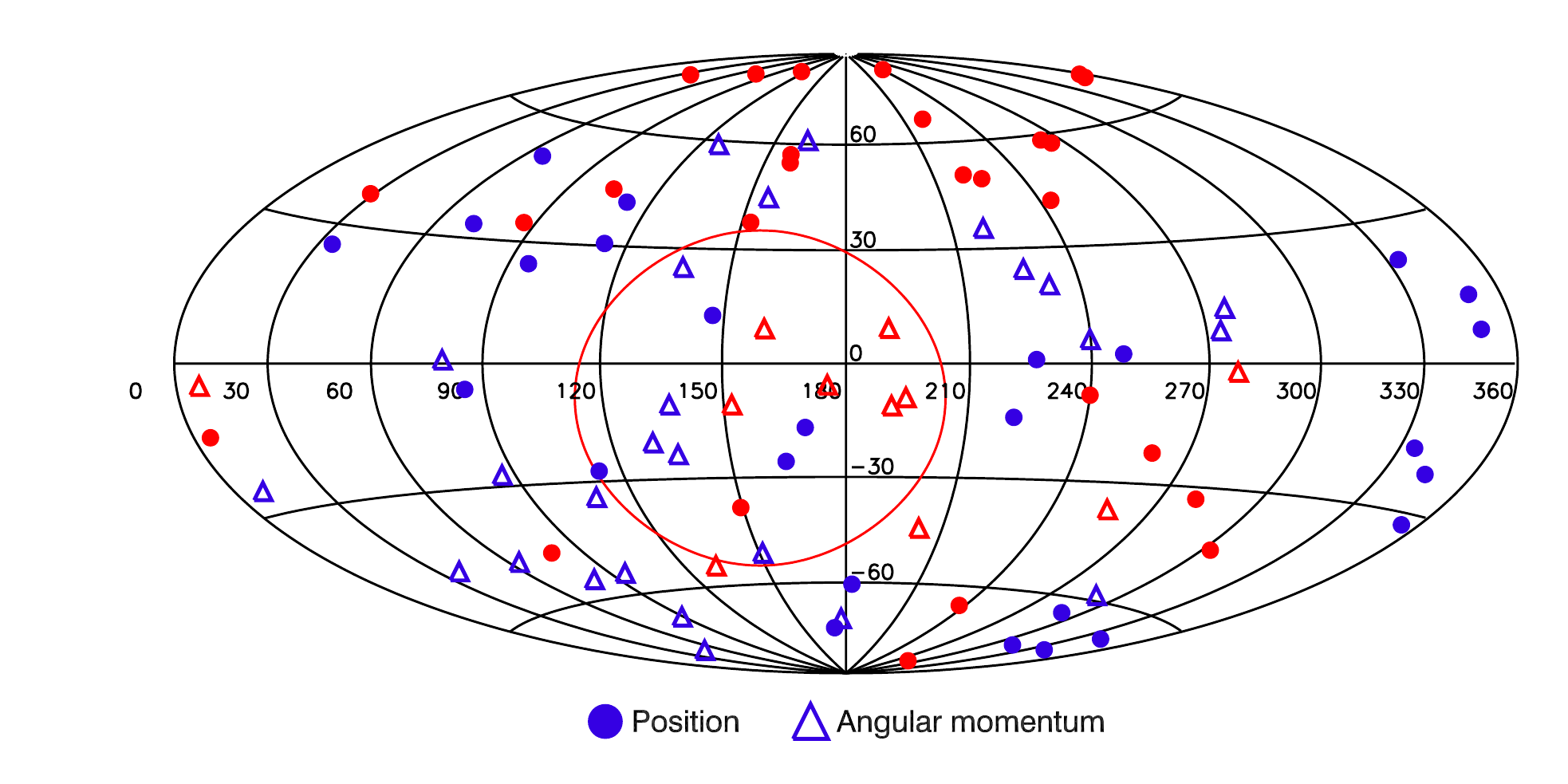}
\caption{
Positions and angular
momenta of the observed (red symbols) satellites and simulated dwarfs (blue symbols) in Galactic coordinates are projected onto an Aitoff Hammer sphere.
The filled circles and triangles represent the positions and angular momenta of the satellites,
respectively. Satellites can be considered as corotating when their angular momenta are clustered in the same direction.
For the observed momenta of 11 satellites, only 6 dwarfs near the center can be considered co-rotating; they are also within 45 degree of the DoS normal
(depicted by red circle). Among the simulated momenta, no strong clustering of the majority of dwarfs are observed.}
\label{fig:aitoff}
\end{figure*}

To further explore the coherent motion of the DoS, we examine the locations of the 18 corotating and 19 counter-corotating dwarfs, as shown in 
Figure~\ref{fig:vert_rad_dist}. Not surprisingly, there is no correlation between spatial distribution and kinematic orientation of
the satellites, the co-rotating or the counter-coorbiting dwarfs are not grouped in either radial or vertical direction,
rather have a random distribution.
This can be further demonstrated in the position -- angular momentum distribution of
the satellites. In Figure~\ref{fig:aitoff}, we compare the Aitoff-Hammer projection of position and angular momentum of the 27 most massive
satellites within the virial radius from the simulation with 27 observed dwarfs of the MW (though only 11 have angular momentum data). Although
6  observed dwarfs (Draco, Umi, SMC, Fornax, Leo II and LMC) may appear to be clustering in the angular momentum distribution (within 45 degree
of 180 degree longitude and 0 degree latitude), spatially they are located far apart in different longitude--latitude planes. Similarly, no strong
angular momentum clustering, or position -- angular momentum correlation, is seen in the simulation. These results suggest that the DoS does not
have coherent rotation.

Similar conclusions have been reported by other studies on DoS. \cite{Cautun2015a} and \cite{Phillips2015} found that the apparent excess of corotating
dwarfs around SDSS galaxies, as claimed by \cite{Ibata2014sdss}, is highly sensitive to sample selection criterion and sample size, and it is 
consistent with the noise expected from an under-sampled data. These findings suggest that the DoS in general is not a kinematically coherent structure.

\section{Evolution of Satellites}
\label{sec:evolution}

\subsection{Evolution of Spatial Distribution}

In order to directly probe the origin of the DoS or the anisotropic distribution of the satellite system, 
it is essential to observe them at high redshift.
However, due to the low luminosity of the satellite galaxies, it is extremely difficult to observe them in the distant universe.
Other than the satellites in our Local Group, we have observations of possible DoS around SDSS galaxies only up to $z=0.05$ \citep{Ibata2014sdss}.
However these claims have been largely refuted \citep{Cautun2015b, Phillips2015}. 

\begin{figure*}
\includegraphics[width=0.3\textwidth, clip=true,trim = 60pt 80pt 130pt 100pt ]{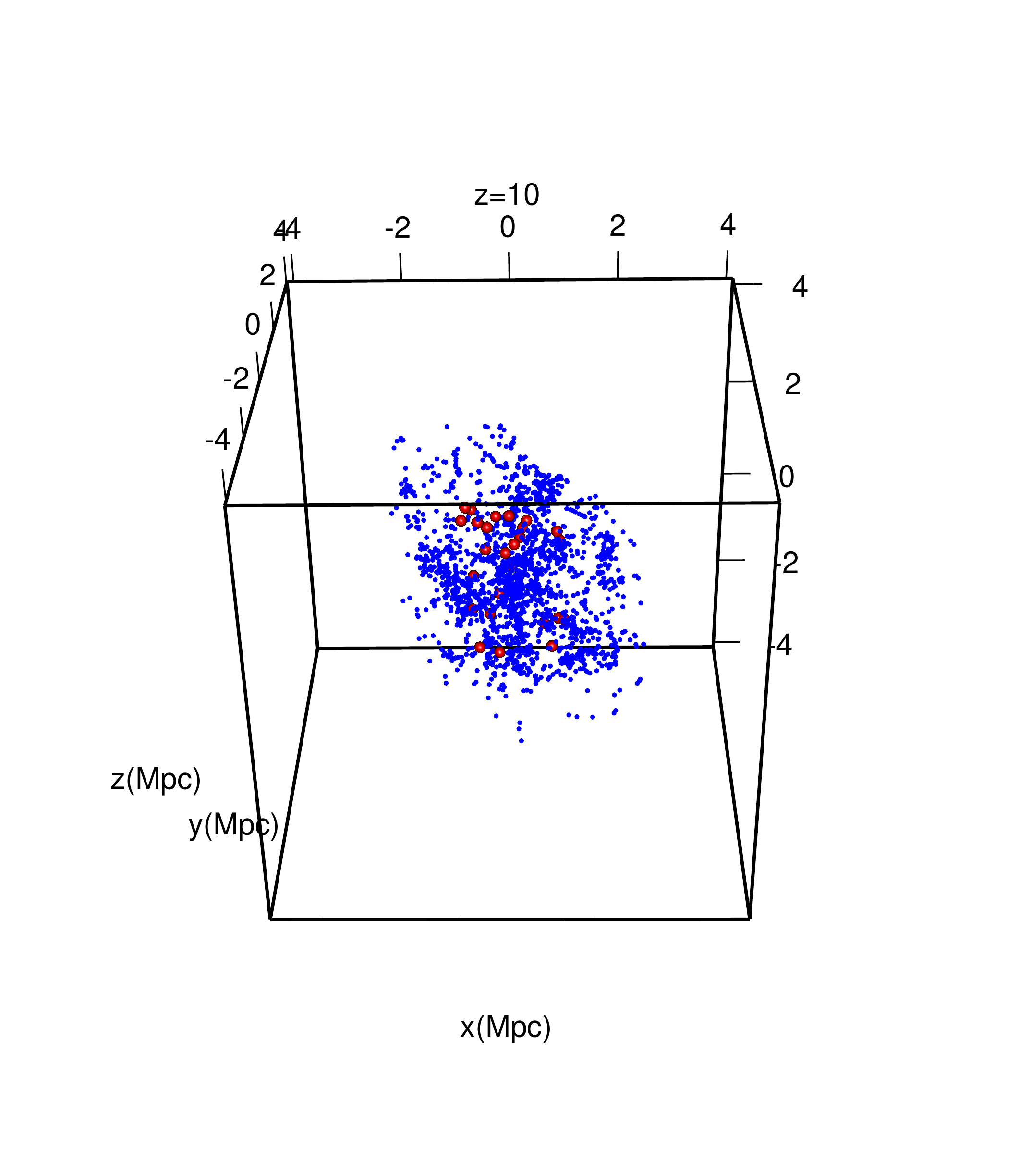}
\includegraphics[width=0.3\textwidth, clip=true,trim = 60pt 80pt 130pt 100pt]{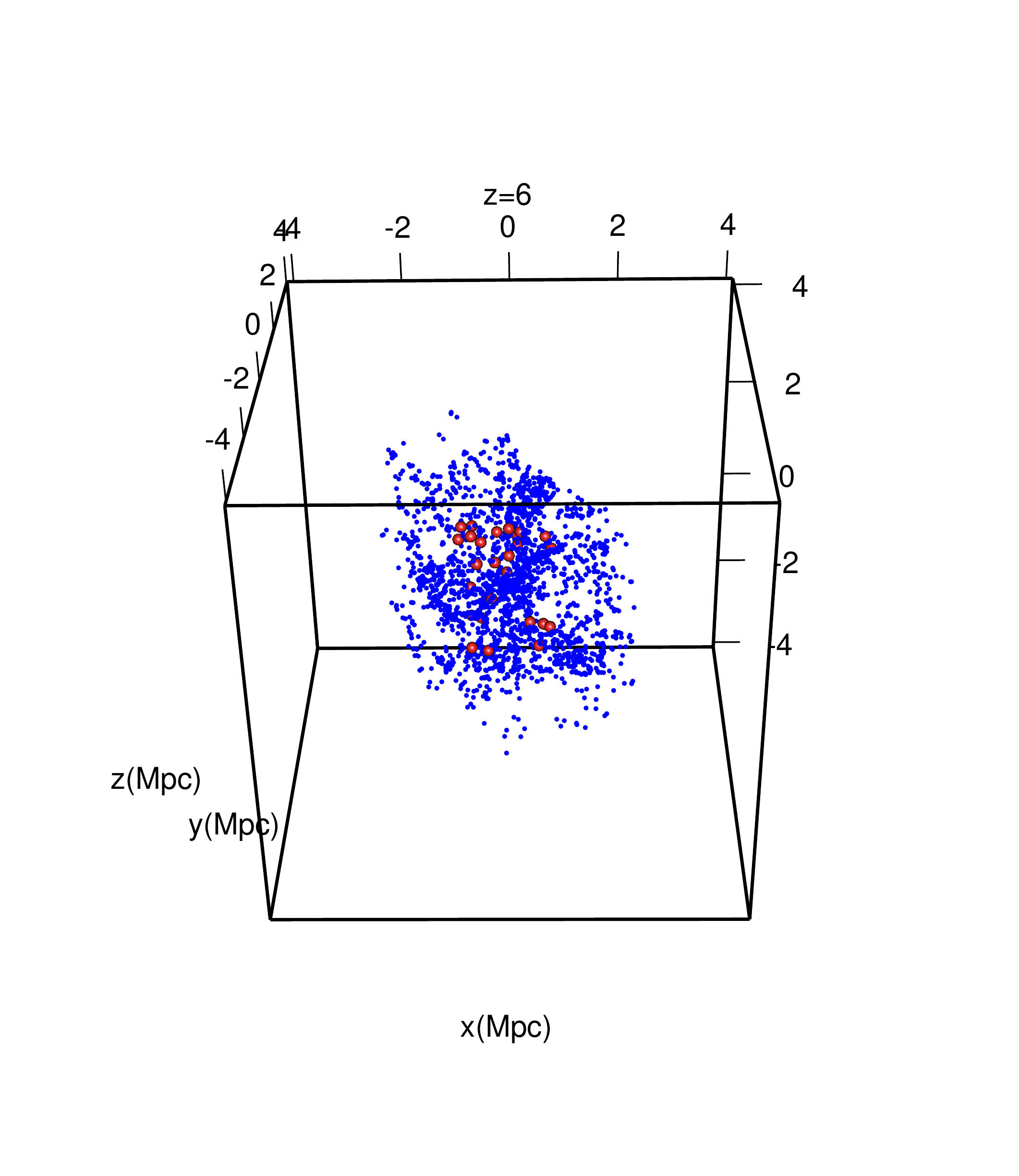}
\includegraphics[width=0.3\textwidth, clip=true,trim = 60pt 80pt 130pt 100pt]{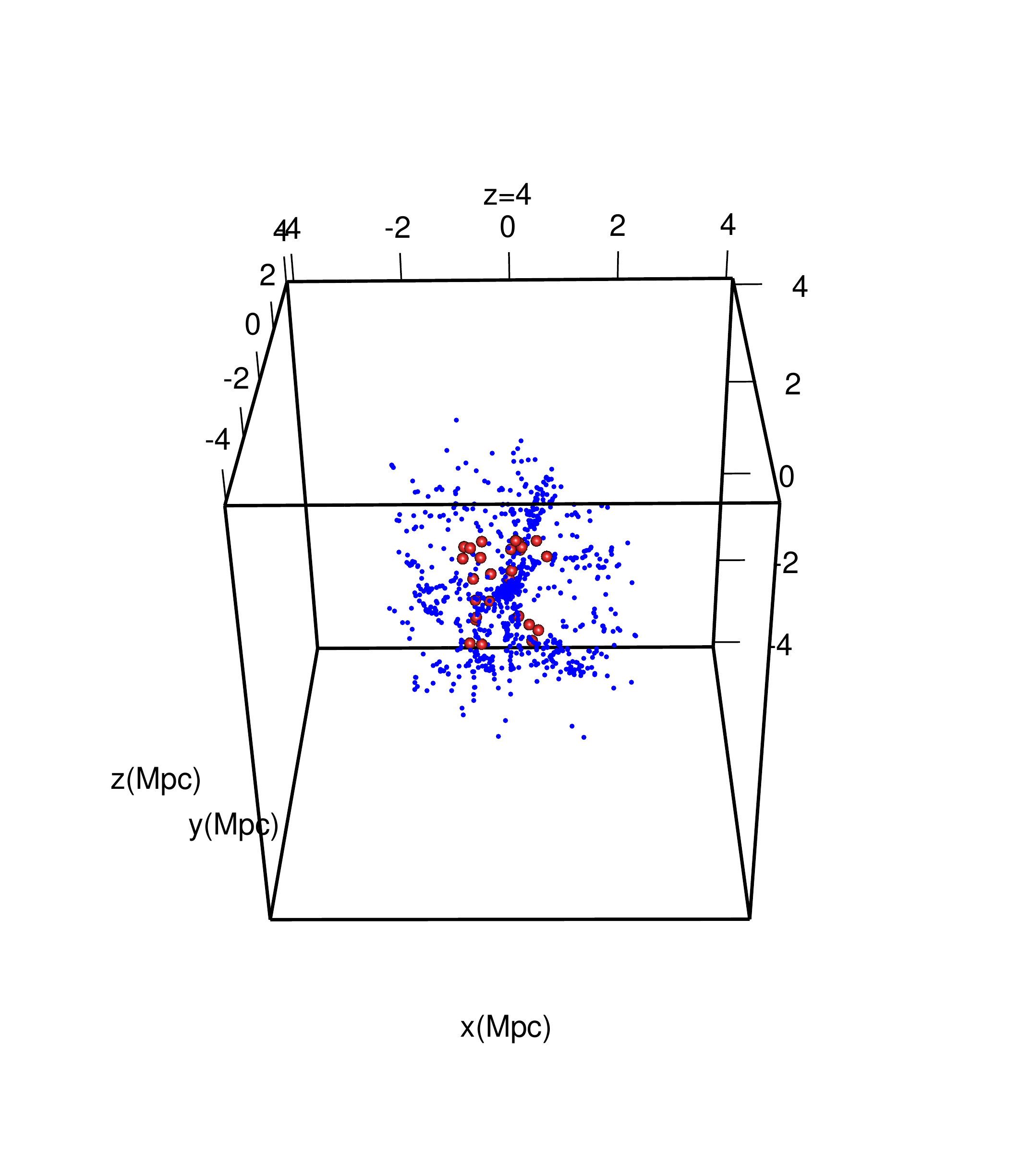}\\
\includegraphics[width=0.3\textwidth, clip=true,trim = 60pt 80pt 130pt 100pt]{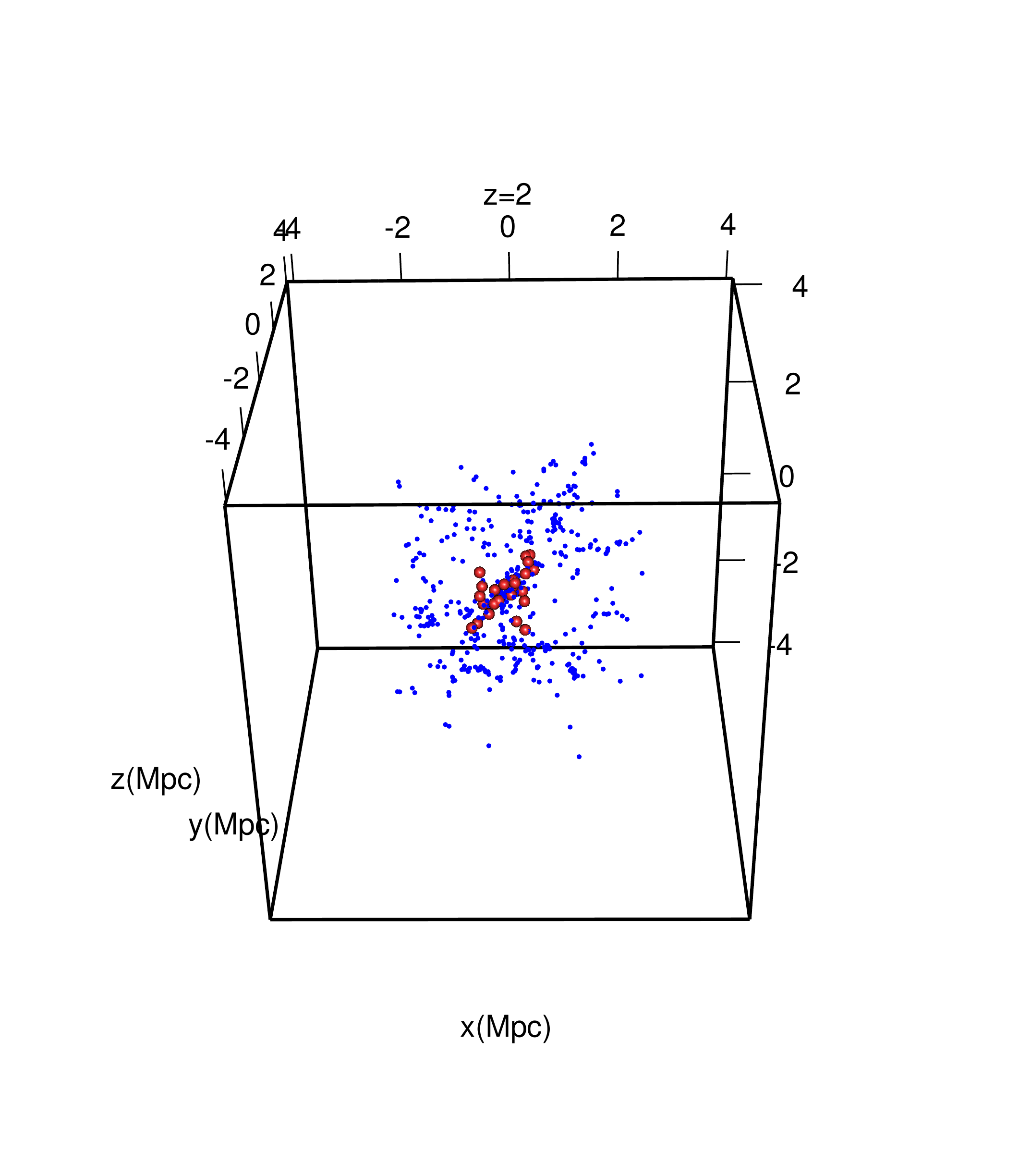}
\includegraphics[width=0.3\textwidth, clip=true,trim = 60pt 80pt 130pt 100pt]{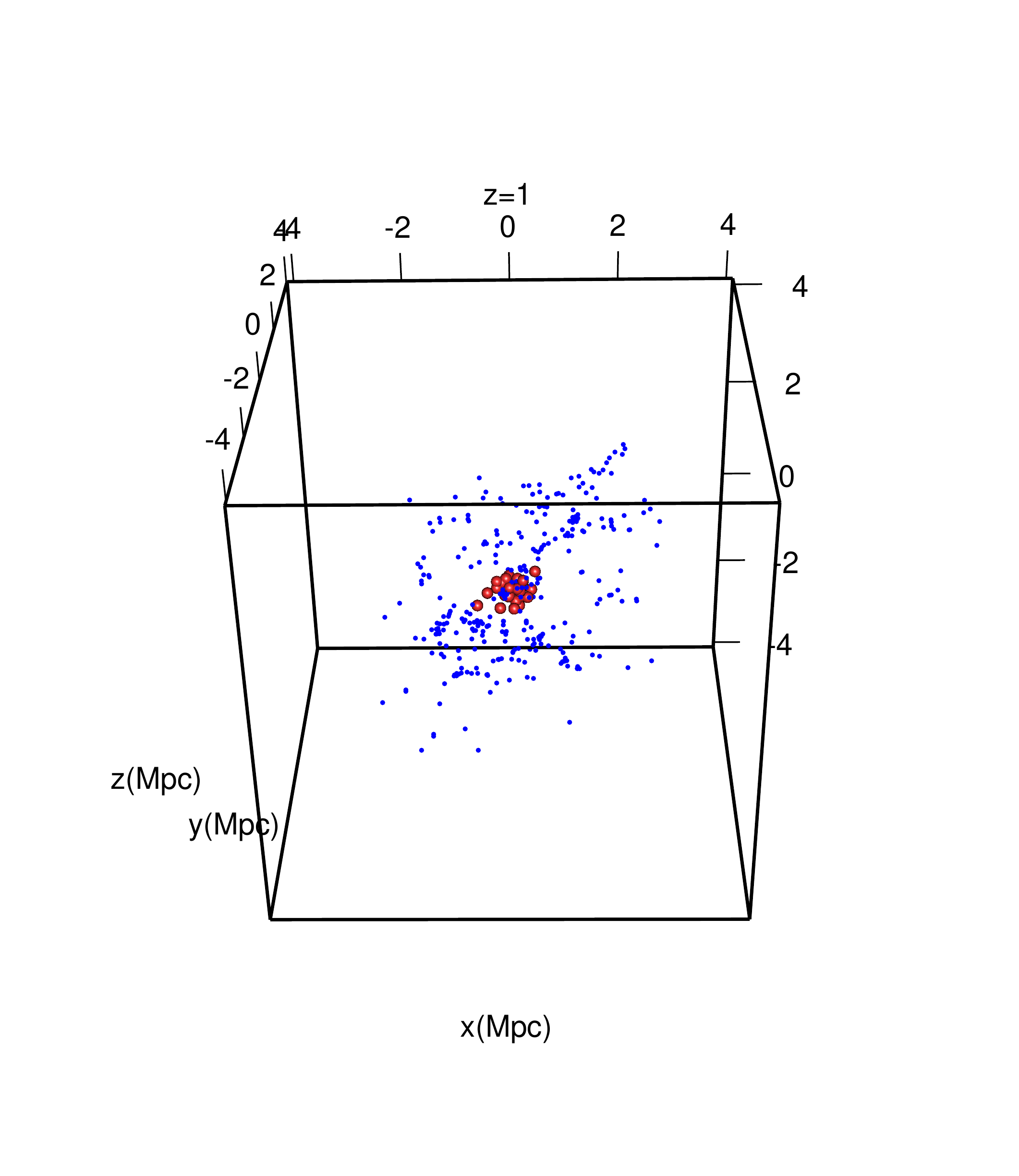}
\includegraphics[width=0.3\textwidth, clip=true,trim = 60pt 80pt 130pt 100pt]{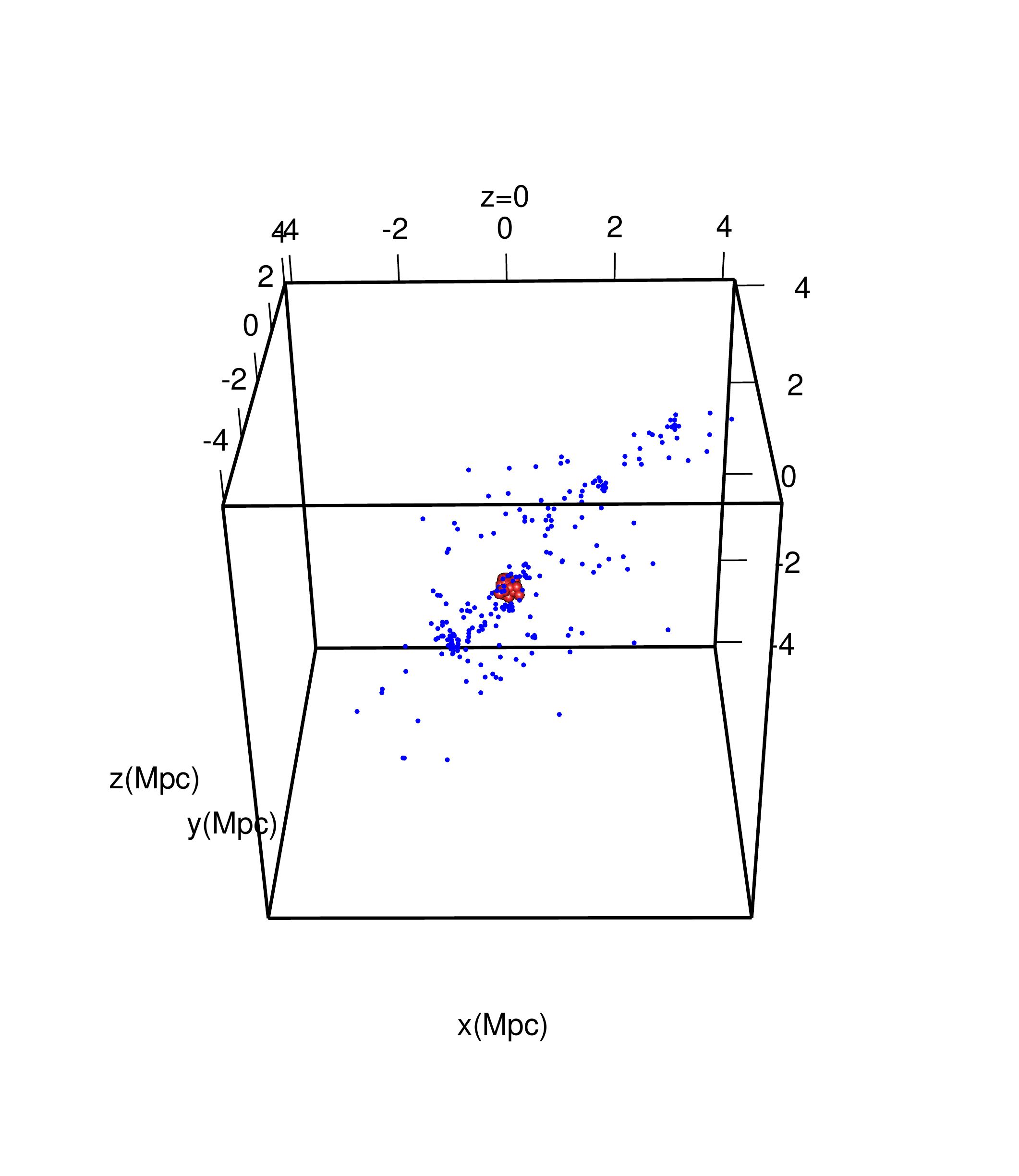}
\caption{Spatial distribution of all dwarfs (blue points) within 3 Mpc of Milky Way (comoving scale) at different redshifts, 
namely $z =10, 6, 4, 1$ and $0$.
The red points are the 27 most massive satellites within 257.4 kpc of the galactic center at $z=0$ which were tracked at higher redshifts. 
The overall distribution of the dwarfs is nearly isotropic at high redshifts but it gradually evolves to be anisotropic with time.}
\label{fig:dwarf_pos}
\end{figure*}

In the simulations we can track the satellite systems to very high redshift. In Fig~\ref{fig:dwarf_pos}, we follow the 3D distribution of the satellites
from redshift $z=10$ to the present day, with the 27 most massive ones highlighted in order to understand those observed in the MW.
At redshift $z=10$, the overall distribution of the galaxies is almost isotropic.
After $z = 6$, the number of dwarfs decreases significantly, 
mostly due the disruption of low-mass halos after reionization \citep{Zhu2016}.
After $z=4$, the galaxies become more strongly clustered along the filaments, as expected from the standard hierarchical structure formation 
model \citep{springel2005millenium, illustris2014}, and the distribution becomes more anisotropic with time. As it approaches to $z=0$,
the distribution of the satellite system becomes strongly anisotropic.

\begin{figure}
\includegraphics[width=0.49\textwidth]{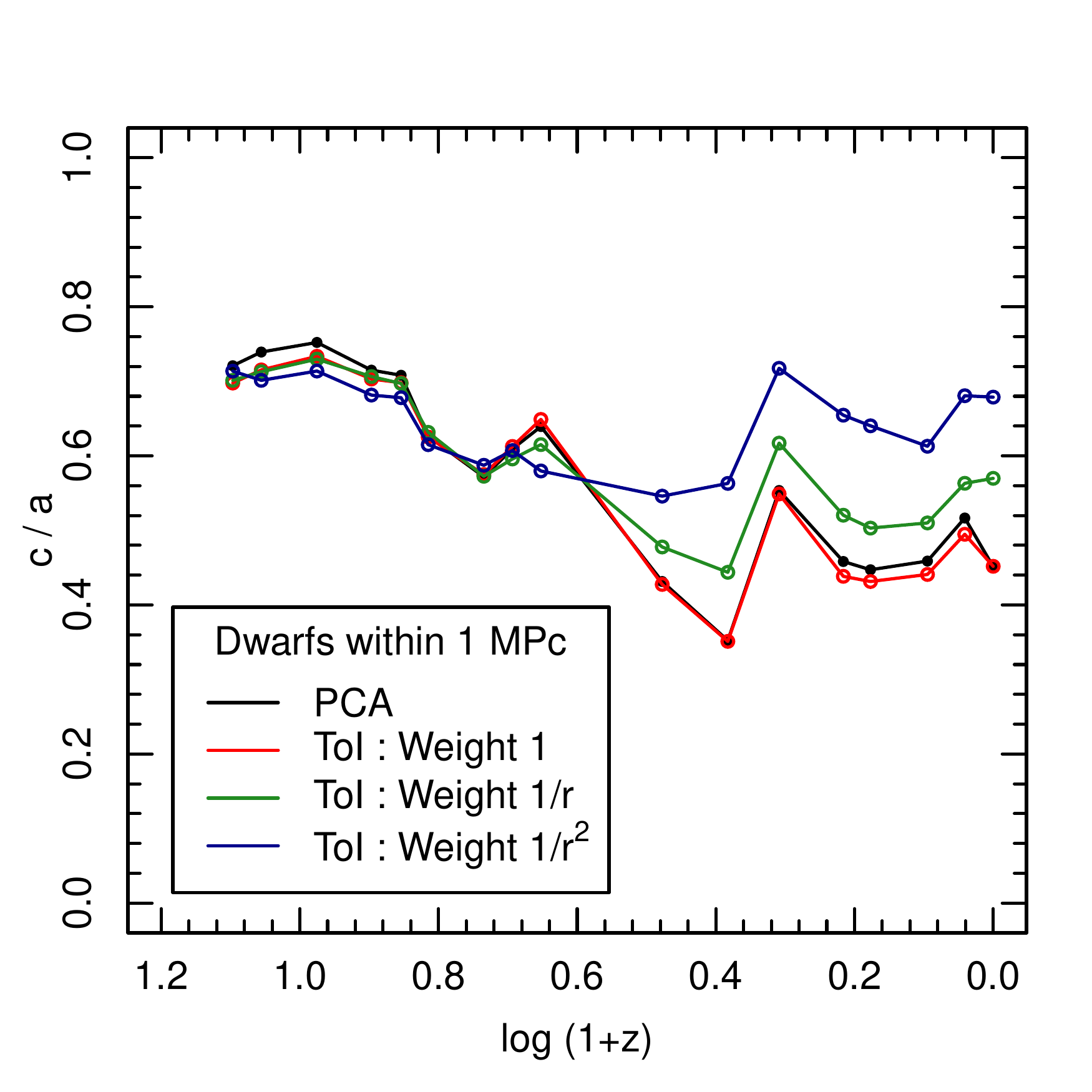}
\caption{Evolution of the isotropy ratio ($c/a$) of the simulated satellite distribution within 1 Mpc.
Different colors the figure denote different plane fitting methods : Principal Component Analysis (black) and
Tensor of Inertia with three types of weight functions, 1 (red), $1/r$ (green) and $1/r^2$ (blue).}
\label{fig:ca_1mpc_bar_evo}
\end{figure}

To quantify the change over time, we fit the satellite system at different redshifts with 4 different plane identification methods
(discussed in section~\ref{sec:4methods}). The resulting evolution of the $c/a$ ratio is shown in Figure~\ref{fig:ca_1mpc_bar_evo}.
We find that all methods give a consistent high $c/a$ ratio at high redshift,
which generally decreases with time. This change is most prominent with both PCA
and the unweighted ToI method, where the ``isotropy'' index starts at $c/a \sim 0.7$ at $ z = 10$ but drops to $c/a \sim 0.44$ at $z = 0$.
These results suggest that the DoS structure is a result of the galaxy formation and evolution process and part of the large-scale filamentary 
structure.

\begin{figure}
 \includegraphics[width=0.49\textwidth]{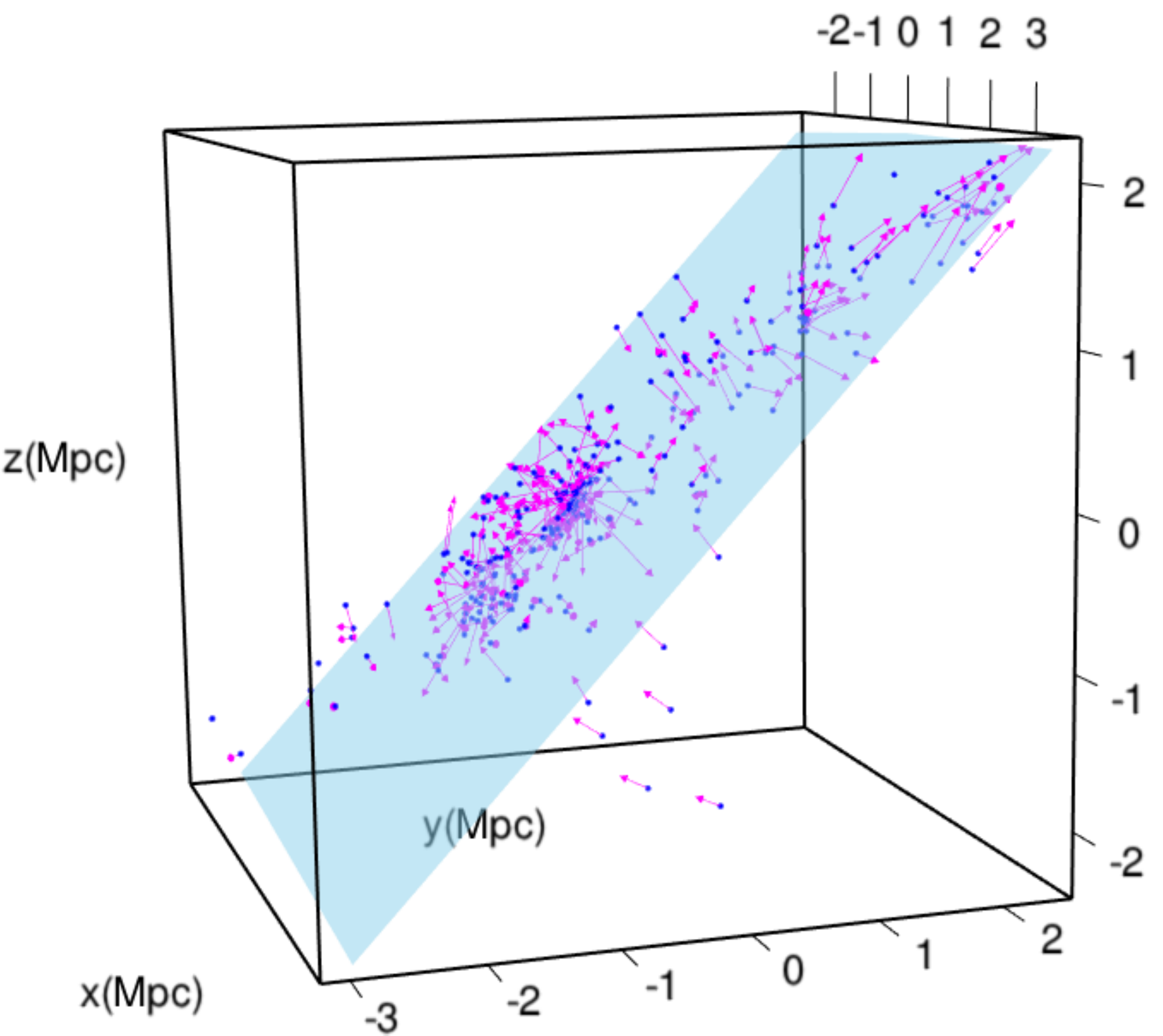}
\caption{Three dimensional plot of simulated dwarf positions (blue points), within 3 Mpc of the galactic center, along with the DoS plane
(blue plane) fitted by PCA method.
The 3D velocities of these dwarf galaxies are represented by the magenta arrows and the length of these arrows is proportional to the 
velocity magnitude.}
 \label{fig:velocity_vectors_3d}
\end{figure}

To investigate the large scale dynamics of the dwarfs, we show their 3D velocity vectors in Figure~\ref{fig:velocity_vectors_3d}.
Around the galactic center, the velocity vectors appear to be random and do not show any preferential rotation, as we have also seen in previous 
sections. This is the result of different accretion history and trajectory of individual dwarfs into the main galaxy 
as shown in \cite{Zhu2016}.
However, at large distances the dwarfs are moving toward a narrow elongated direction, which suggests that they are part
of the large scale filamentary structure.

\subsection{Evolution of the Kinematics}

\begin{figure}
\includegraphics[width=0.49\textwidth]{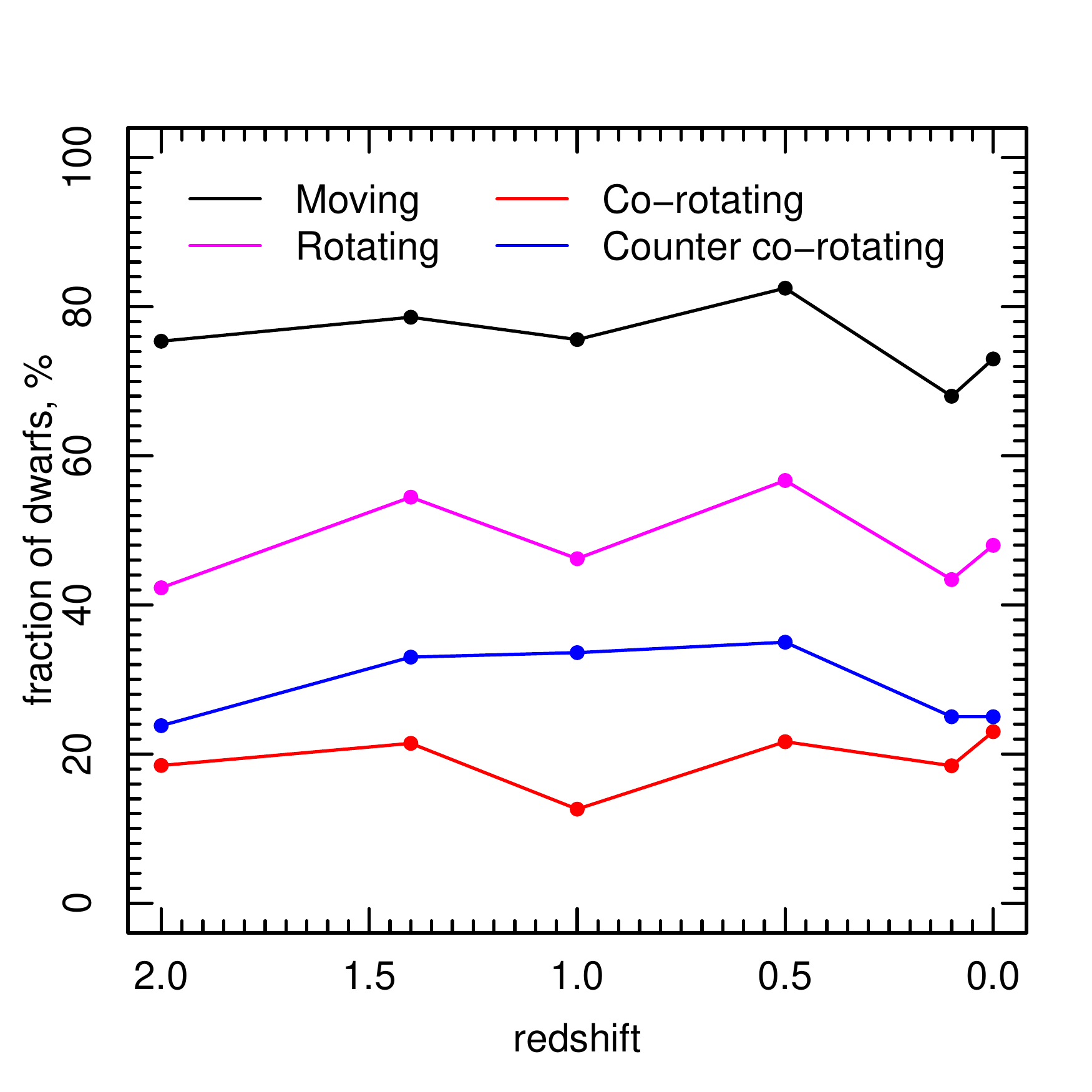}
 \caption{Evolution of the fraction of dwarfs (within 1 Mpc of galactic center) with different kinematical properties : 
the fraction of dwarfs moving on the DoS plane (black), the fraction of dwarfs co-rotating in DoS (red), 
the fraction of dwarfs counter co-rotating in DoS (blue), and the total fraction of dwarfs rotating (corotating and counter-corotating)
in DoS (magenta).}
\label{fig:kinematic_evolution}
\end{figure}

The evolution of the kinematic properties of the dwarfs is shown in Figure~\ref{fig:kinematic_evolution}, in which
we track dwarf subsets with different kinematical properties: dwarfs moving on the DoS and
rotating ones which include corotating and counter-corotating. 
We find that over the redshift range from $z = 2$ to $z = 0$, the different fractions remain nearly the same. Around $80\%$ of these galaxies
are primarily moving on DoS (as opposed to normal to DoS) and around $45\%$ of them are rotating on the DoS, the rest is moving
radially.

Interestingly, the fractions of corotating and counter-corotating dwarfs are comparable at around $20\%$ throughout the time. 
This reaffirms our conclusion that the DoS has no coherent rotation and it is not rotationally supported.

\section{Discussions}
\label{sec:discussions}



\begin{figure}
\includegraphics[width=0.49\textwidth]{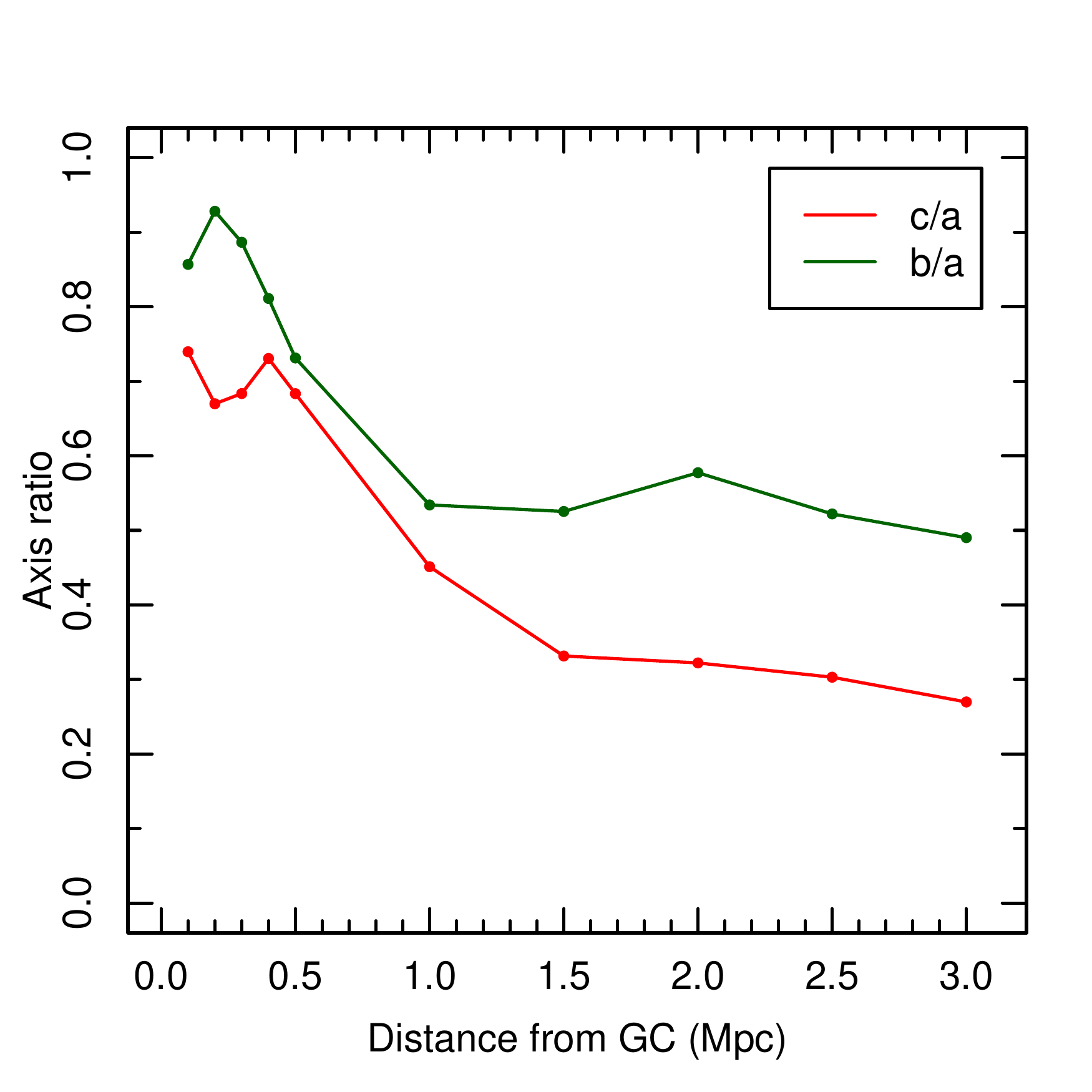}
\caption{The $c/a$ ratio (red) and the $b/a$ ratio (green) of the dwarf distribution as a function of the maximum distance of the 
dwarfs from galactic
center. For each distance, we take all dwarfs within that radius and calculate these two ratios using PCA method.}
\label{fig:ca_ba}
\end{figure}

In order to investigate how the DoS structure changes with the distance from the central galaxy, we plot 
the two different axis ratios ($c/a$ and $b/a$) of the dwarf distribution at
different radii from the galactic center in Figure~\ref{fig:ca_ba}.
We find that both ratios decreases with increasing distance, 
by 3 Mpc, 
$c/a \sim 0.25$ and $b/a \sim 0.5$, which may resemble the large-scale filamentary structure. 
However, we note that
when we include all the dwarfs (not just the most massive ones as in the previous sections),
the distribution becomes more isotropic close to the galactic center compared to the distribution of only massive dwarfs.

\begin{figure*}
\centering
 \includegraphics[width=0.49\textwidth]{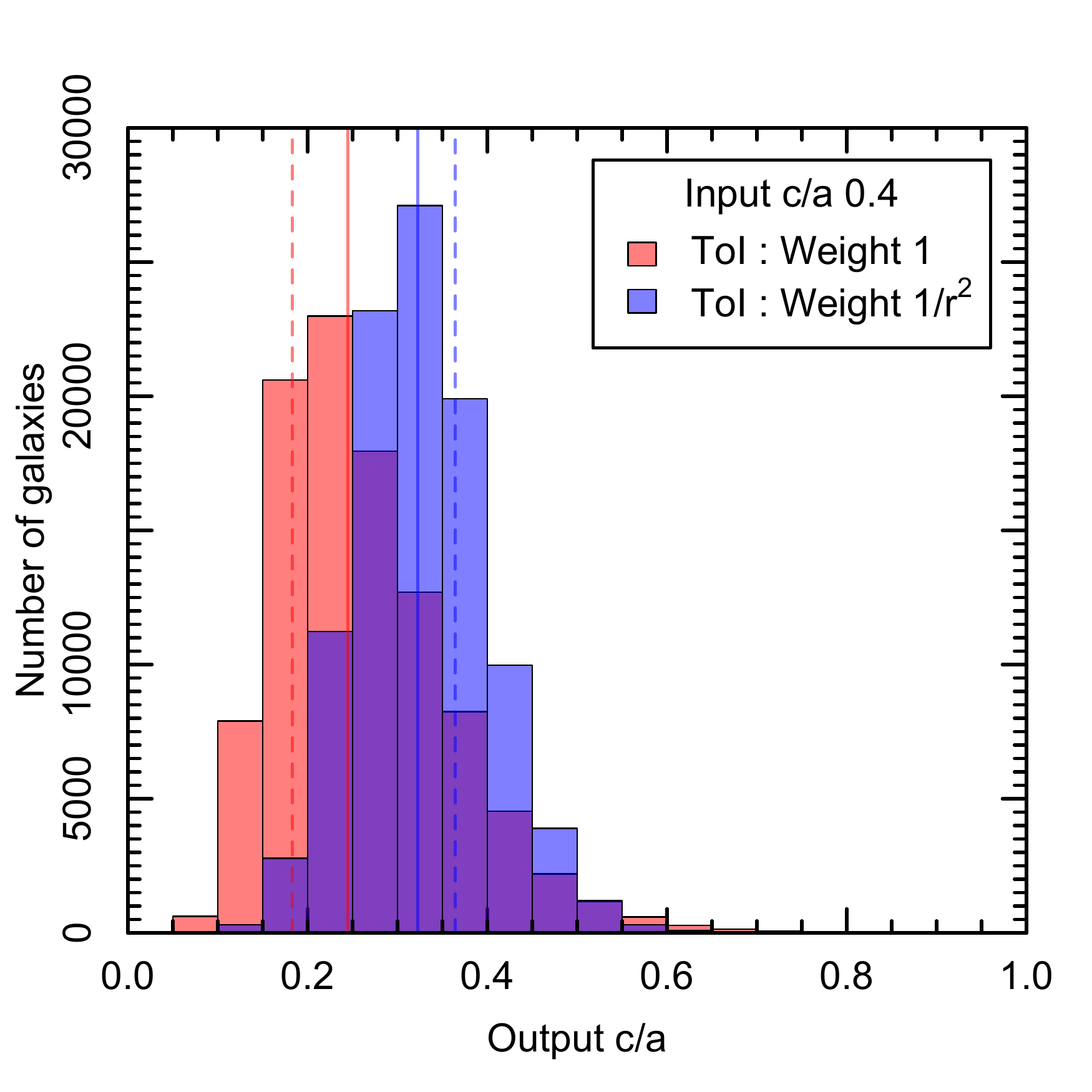}
  \includegraphics[width=0.49\textwidth]{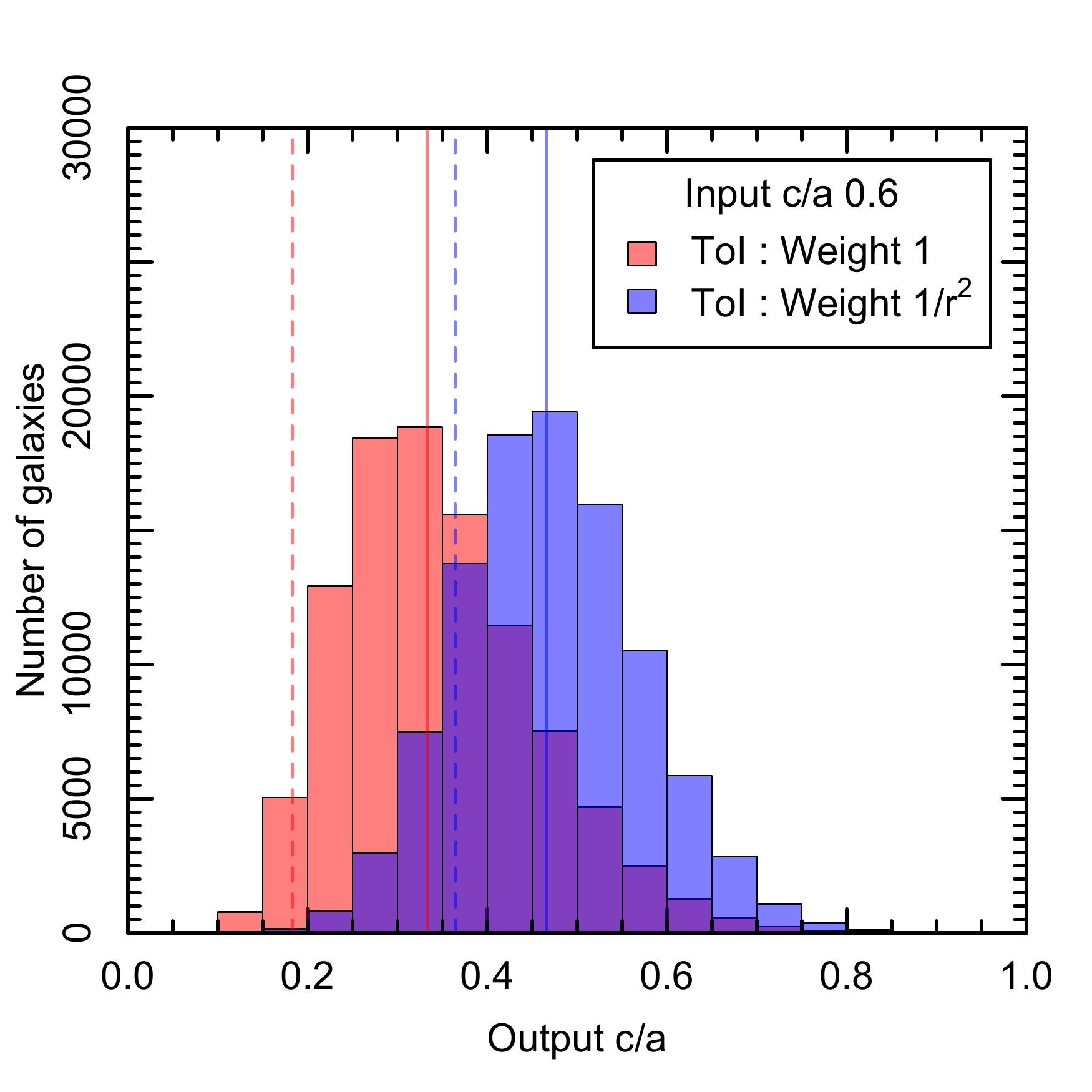} \\
   \includegraphics[width=0.49\textwidth]{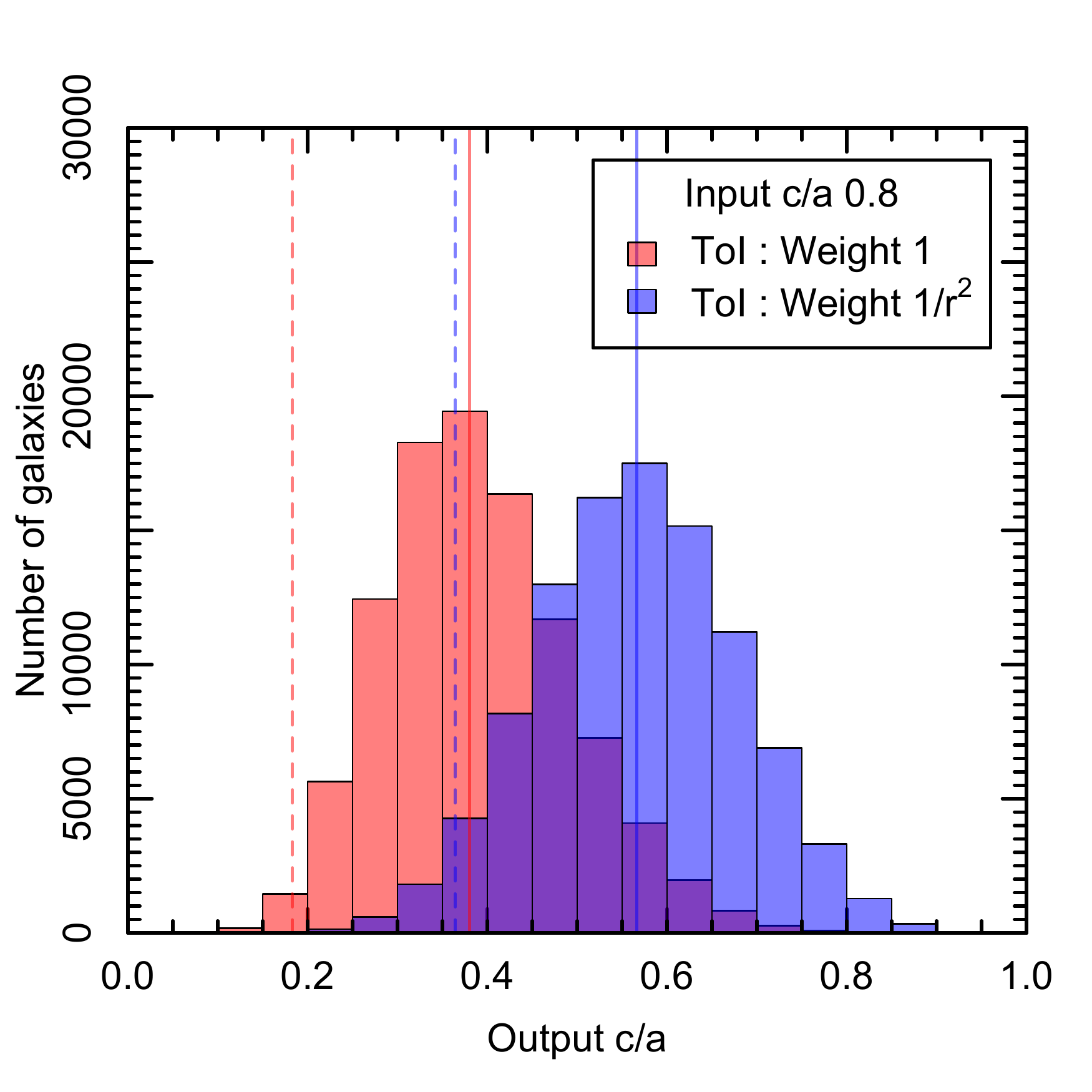}
    \includegraphics[width=0.49\textwidth]{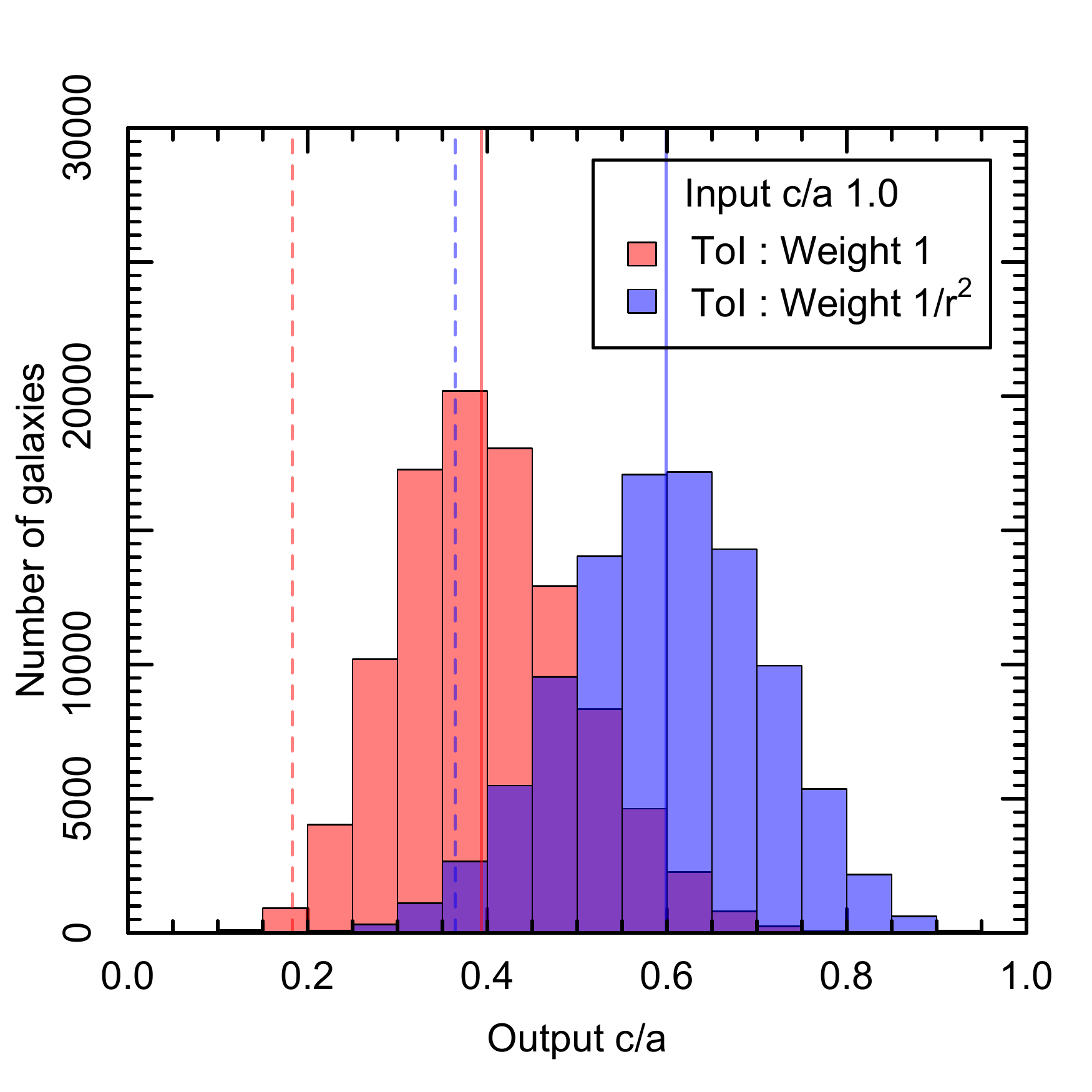}
 \caption{Distribution of output galaxy c/a for different input values ($c/a = 0.4$ top left panel, 0.6 in top right panel,
 0.8 in bottom left panel and 1.0 in bottom right panel) in a Monte Carlo simulation with 100,000 galaxies. The c/a ratio is calculated by two methods:
weighted by 1 (red), and weighted by $1/r^2$ (blue). The violet region shows the overlap between the distributions with two methods.
The dashed vertical lines in red and blue shows the observed c/a value calculated with
11 MW satellites with these two methods respectively. The median $c/a$ values for the systems are shown with vertical red (weight 1)
and blue (weight $1/r^2$) solid lines.}
\label{fig:MonteCarlo}
\end{figure*}

Furthermore, we note that the $c/a$ ratio may not adequately represent the underlying distribution of the system.
To demonstrate this, 
we take the observed positions of the 11 classical dwarfs and perform four tests on them. 
In the first test, we distribute the 11 dwarfs in their observed distances but 
placing them in a way such that their input $c/a = 0.4$.
Now we perform a Monte Carlo simulation for 100,000 realizations of this system and calculate the $c/a$ ratio of each realization
using two weighted methods (weights 1, and $1/r^2$ respectively, details in \S 2.3.2). 
We plot the distribution of the output $c/a$ ratio of these galaxies in Figure~\ref{fig:MonteCarlo} (top left panel) and find that although the 
input $c/a = 0.4$, about $\sim 20\%$ of the systems have an output $c/a \lesssim 0.18$, the observed anisotropy ratio for the observed 11 galaxies
(with weight 1 method). We repeat this test with three other input $c/a$, 0.6, 0.8 and 1.0
respectively and find that for each of them there is non-negligible probability that the system has a lower output $c/a$ than the input value. 
We also notice that the method used to determine the $c/a$ ratio also influences the output, for all samples, the weighted by $1/r^2$ method
produces higher $c/a$ than the unweighted method. This shows that very small samples, i.e. 11 dwarfs, may artificially indicate a higher anisotropy
and they may not contain the full information of the underlying distribution.


Finally we stress that our results are subject to the limitations of our study, as we have only 
analyzed one particular realization of a MW-sized galaxy. In future projects we plan to pursue a more systematic study by extending 
our simulated galaxy sample. 

\section{Summary}
\label{sec:summary}

We have investigated the spatial distribution and kinematic properties of satellites of a MW-sized galaxy by comparing a 
high resolution hydrodynamical cosmological simulation with its DM only counterpart.
Our main results are summarized as follows:

\begin{itemize}
\item Baryons play a significant role in determining the abundance and distribution of the satellite system of a galaxy.
Within 1 Mpc from the central galaxy, only 106 dwarf galaxies containing stars are found at the present day in the 
hydrodynamic simulation and they show an anisotropic
distribution, in sharp contrast to 21,220 subhalos in the N-body 
simulation, which are distributed isotropically;

\item The DoS in our simulation is not rotationally supported and there is no coherent motion of the satellites,
as the fraction of corotating and counter-corotating satellites are 
comparable and around $19\%$ across cosmic time.

\item 
The distribution of the (baryonic)
satellite galaxy system evolves significantly with time. It is highly isotropic at high redshifts but it becomes more 
anisotropic as redshift approaches to $z=0$, 
and this anisotropic distribution is part of the 
filamentary structure in the hierarchical structure formation scenario.

\item The properties of the DoS strongly depend on the sample size and the plane identification methods.
When only the 11 most massive dwarfs similar to those ``classical'' Milky Way satellites are selected,
the DoS becomes more flattened as observed. 
However, when the sample size increases the DoS becomes thicker. This is consistent with the observational
 pattern that height and the flatness ratio of the DoS increase with sample size, as shown in Maji et. al. 2017 (in prep.).

\end{itemize}

Our results suggest that the highly-flattened, coherently-rotating DoS claimed in the MW and other galaxies may be a selection effect 
due to a small
sample size, and that 
different subhalo distributions that we see in our hydrodynamical and N-body simulations are shaped by baryons.
Baryonic processes such as adiabatic contraction, reionization, and tidal destruction of galaxies can have significant effects on the abundance,
star formation, infall time and trajectories of the satellites which can in turn affect their final distribution. Therefore, effects of baryons 
should be taken into account in the study of the distribution and evolution of the satellite system of a galaxy.

\bibliography{dos}

\end{document}